\documentclass[]{aa}
\usepackage{graphicx}
\usepackage{epsfig}
\usepackage{subfigure}
\usepackage{amssymb}
\usepackage{amsmath}
\usepackage{textgreek}
\usepackage{natbib}
\bibpunct{(}{)}{;}{a}{}{,}

\newcommand{\FIG}[1]{#1}

\def\mso{\,{\rm M}_\odot}

 \def\rso{\,{\rm R}_\odot}

 \def\gcm{\,{\rm g}\,{\rm cm}^{-3}}
 
 \def\kms{\, {\rm km}\, {\rm s}^{-1}}

 \def\simle{\mathrel{\hbox{\rlap{\hbox{\lower4pt\hbox{$\sim$}}}\hbox{$<$}}}}
 \def\simgr{\mathrel{\hbox{\rlap{\hbox{\lower4pt\hbox{$\sim$}}}\hbox{$>$}}}}
 \def\vinf{\, \mathrm{v}_\infty}
 
 \def\mdot{\, \dot{M}}
 \def\msoy{\, \mso~{\rm yr}^{-1}}
 \def\muG{\,\textmu G}

\def\pramism{\, P_{\rm ram}}
\def\pramwind{\, p_{\rm ram}}
\def\ptism{\, P_{\rm T}}
\def\ptwind{\, p_{\rm T}}
\def\pbism{\, P_{\rm B}}
\def\pb{\, P_{\rm B}}

\def\rb{\, R_{\rm B}}
\def\mh{\, m_{\rm h}}
\def\vel{\, \mathbf v}
\def\Bfield{\, \mathbf B}

\begin{document}

   \title{Shape and evolution of wind-blown bubbles of massive stars: \\ on the effect of the interstellar magnetic field}

   \author{A. J. van Marle
          \inst{1,2}
          \and
          Z. Meliani
          \inst{3,4}
          \and
          A. Marcowith
          \inst{5}
          }

   \offprints{A. J. van Marle}

   \institute{
              KU Leuven, Centre for mathematical Plasma Astrophysics, 
              Celestijnenlaan 200B, B-3001 Leuven, Belgium \\
              \email{AllardJan.vanMarle@wis.kuleuven.be}
       \and
              KU Leuven, Institute of Astronomy, 
              Celestijnenlaan 200D, B-3001 Leuven, Belgium 
       \and
            LuTh, Observatoire de Paris, 
            5 place Jules Janssen, 92195 Meudon, France \\
              \email{zakaria.meliani@obspm.fr}
       \and 
           APC, Universit{\'e} Paris Diderot, 
           10 rue Alice Domon et L{\'e}onie Duquet, 75205 Paris Cedex 13, France
       \and
            Laboratoire Univers et Particules (LUPM) Universit{\'e} Montpellier, CNRS/IN2P3, CC72,
            place Eug{\`e}ne Bataillon, F-34095 Montpellier Cedex 5, France \\
            \email{Alexandre.Marcowith@univ-montp2.fr}
}

\date{Received <date> / Accepted <date>}

\abstract
{The winds of massive stars create large ($>10$\,pc) bubbles around their progenitors. 
As these bubbles expand they encounter the interstellar coherent magnetic field which, depending on its strength, 
can influence the shape of the bubble. }
{We wish to investigate if, and how much, the interstellar magnetic field can contribute to the shape of an expanding circumstellar bubble 
around a massive star.}
{We use the \emph{MPI-AMRVAC} code to make magneto-hydrodynamical simulations of bubbles, using a single star model, 
combined with several different field strengths: $B\,=\,5,\,10,$\,and\,$20$\,\muG\, for the interstellar magnetic field. 
This covers the typical field strengths of the interstellar magnetic fields found in the galactic disk and bulge.
Furthermore, we present two simulations that include both a $5\,$\,\muG\, interstellar magnetic field and a warm (10\,000\, K) interstellar medium (ISM) 
and two different ISM densities  
to demonstrate how the  magnetic field can combine with other external factors to influnece the morphology of the circumstellar bubbles.}
{Our results show that low magnetic fields, as found in the galactic disk, inhibit the growth of the circumstellar bubbles in the direction
perpendicular to the field. As a result, the bubbles become ovoid, rather than spherical. 
Strong interstellar fields, such as observed for the galactic bulge, 
can completely stop the expansion of the bubble in the direction perpendicular to the field, leading to the formation of a tube-like bubble. 
 When combined with an ISM that is both warm and high density the bubble is greatly reduced in size, causing a dramatic change in the evolution 
of temporary features inside the bubble such as Wolf-Rayet ring nebulae.}
{The magnetic field of the interstellar medium can affect the shape of circumstellar bubbles. 
This effect may have consequences for the shape and evolution of circumstellar nebulae and supernova remnants, 
which are formed within the main wind-blown bubble.}

  \titlerunning{Magnetic ISM and circumstellar bubbles}
  \authorrunning{van Marle, Meliani \& Marcowith}


  \maketitle

%

\section{Introduction}
Massive stars lose a significant fraction of their mass in the form of stellar wind, which expands into the surrounding medium. 
As it expands, the stellar wind collides with the gas in the interstellar medium (ISM), creating a low density bubble, expanding over time. 
Analytically, the general shape of such a wind blown bubble was predicted by \citet[e.g][]{Adevisova:1972} and \citet{Weaveretal:1977}. 
However, these models were limited, because they were strictly one-dimensional and could therefore not take into account the effects of either a-spherical stellar winds, 
or irregularities in the ISM. 
Furthermore, they assumed that the stellar wind properties would remain constant over long periods of time. 
These analytical models were complemented by numerical simulations, which can include both a non-uniform ISM and time-dependent wind parameters 
\citep[E.g.][]{Rozyczka:1985,RozyczkaTenorioTagle:1985a,RozyczkaTenorioTagle:1985b}. 
Further developments in both numerical hydrodynamics and stellar evolution theory made it possible to follow the evolution 
of the stellar wind parameters as the progenitor star evolves, and to investigate the effects on the morphology of circumstellar bubbles.

Initial models by \citet{GarciaSeguraetal:1996a, GarciaSeguraetal:1996b} 
focussed primarily on the transitional phases of the stellar evolution, 
e.g. the transition from main sequence to either a Luminous Blue Variable or a Red Supergiant and the transition from 
the giant stage to the Wolf-Rayet stage. 
These models proved that the radical changes in stellar wind parameters accompanying  
such transitions lead to the formation of shells that can be observed as circumstellar nebulae. 
Subsequent models by \citet{Freyeretal:2003, Freyeretal:2006} and \citet{vanmarleetal:2005, vanmarleetal:2007,vanmarleetal:2008}
followed the entire evolution of the star and introduced the effect of photo-ionization. 
\citet{TenorioTagleetal:1990} and \citet{Dwarkadas:2005,Dwarkadas:2007} added the supernova explosion and followed the evolution of the supernova remnant inside the circumstellar bubble. 
Recently, \citet{ToalaArthur:2011} included radiative transfer to investigate the ionization structure of the circumstellar medium 
and added thermal conduction to their models. 
 \citet{Geenetal:2015} made similar models for a somewhat lower mass star (15\,$\mso$) in 3D.

Another factor potentially affecting the evolution of circumstellar bubbles is the interstellar magnetic field. 
Observations show that the interstellar medium contains large scale coherent magnetic fields.
These fields are usually rather weak \citep[$1\lesssim B\lesssim 60$\muG\, according to e.g.][]{RandKulkarni:1989,OhnoShibata:1993,Fricketal:2001,Opheretal:2009,
Shabalaetal:2010,Crockeretal:2011,Fletcheretal:2011,HeerikhuisenPogorelov:2011,Vallee:2011,Greenetal:2012},  
but can potentially influence the shape of the interstellar bubbles, blown by stellar winds. 
(Higher values have been found for the interior of molecular clouds, though these typically occur in those regions where the cloud is at its densest, 
rendering them relatively small in scale. Even so, \citet{Crutcheretal:1999} shows a field strength of 480\muG\, with a radius of 22\,pc for \object{SgrB2}.) 
Despite their low strength, interstellar magnetic fields can influence the shape of circumstellar bubbles. 
This influence can occur on a relatively small scale, since the presence of a magnetic field changes 
the properties of hydrodynamical instabilities \citep{Junetal:1995,Stonegardiner:2007}.
This effect was described analytically \citep{Dganietal:1996,Breitschwerdtetal:2000} 
and shown in numerical models of a stellar wind bow shock by \citet{vanMarleetal:2014}. 
On a larger scale, the interstellar magnetic field can influence the general shape of a circumstellar bubble by reducing the expansion rate in the direction perpendicular to the magnetic field. 
This was already predicted by \citet{Heiligman:1980} and worked out in detail for the evolution of planetary nebulae by \citet{SokerDgani:1997}. 
More recently, \citet{FalcetaGoncalvesMonteiro:2014} showed numerically that a strong interstellar magnetic field ($\simeq\,500$\muG) could lead to the creation of 
a bipolar planetary nebula. 
A similar effect was noted by \citet{vanMarleetal:2014b}, who used numerical simulations to demonstrate that interstellar magnetic fields can explain the formation 
of ``eye-like'' structures around certain asymptotic giant branch (AGB) stars \citep{Coxetal:2012}.

Hot, massive stars have strong winds, which might allow them to overcome the forces exerted by the magnetic field. 
On a large scale, the influence of the interstellar magnetic field on the shape of super bubbles, which contain clusters of massive stars, 
was investigated numerically by \citet{Tomisaka:1990,Tomisaka:1992} and \citet{Ferriere:1991} 
who found that a parallel magnetic field constrains the expansion of the bubble in the direction perpendicular to the field. 
\citet{Tomisaka:1998} investigated the 3D dynamics of a superbubble in stratified and magnetized medium, exploring the conditions under which the bubble can blow out in the halo. 
However, these models considered supernova explosions to be the only source of energy in the bubble and approximated 
the effect of these sequential supernovae with an input of energy that remained constant in time. 
No attempt was made to include the stellar wind as a separate force to drive the initial expansion of the bubble, or to investigate how the 
deviation from spherical symmetry of the outer shell influences the morphology of the circumstellar shells formed inside the bubble 
when the properties of the stellar wind change or during the expansion of a supernova. 
Although the total energy contribution of the supernova to the circumstellar bubble is approximately three times larger than the wind energy 
contribution (See Table~\ref{tab:starpars}), 
the wind shapes the bubble in which the supernova expands and therefore should not be neglected in the simulations. 
We use the \emph{MPI-AMRVAC} magneto-hydrodynamics code \citep{Keppensetal:2012} to run a series of simulations of circumstellar bubbles, 
using a generic 40$\mso$ star model like the one used in \citet{vanMarleetal:2012a}. 
By varying the strength of the interstellar magnetic field, while keeping all other quantities constant, we investigate both its qualitative and quantitative effects on the wind expansion 
to see if, and how, the magnetic field influences the shape and evolution of the circumstellar bubble. 
Additionally, we present two simulations showing similar expansions in a warm, rather than a cold ISM to demonstrate 
how the interstellar field can combine with other external factors to influence the size and shape of circumstellar bubbles.
Instead of using a constant inflow of energy at the centre, as was done by \citet{Tomisaka:1990,Tomisaka:1992} and \citet{Ferriere:1991}, 
we approximate the stellar evolution with a three-stage model (main sequence, red supergiant, and Wolf-Rayet) followed 
by a supernova explosion. This allows us to follow the evolution of the temporary shells formed inside the bubble 
as a result of changes in the wind parameters and the eventual supernova explosion. 

\subsection{Layout}
In Sect.~\ref{sec-ISM} of this paper we describe the physical considerations that are involved in the interaction between the expanding stellar wind 
and the interstellar magnetic field and use an analytical approximation to predict the field-strength necessary to overcome the ram-pressure of the wind. 
We then describe our model parameters and numerical method in Sect.~\ref{sec-method}. 
The results of our simulations for a cold ISM are shown in Sect.~\ref{sec-results}. The case of a warm interstellar medium is shown in Sect.~\ref{sec-warm}.  
All cases are discussed in Sect.~\ref{sec-discussion}.  
Finally, in Sect.~\ref{sec-conclusions}, we present our conclusions.

For full animations of our results in electronic form, see Appendix~\ref{sec-animations}

 \begin{figure}
\FIG{
 \centering
\mbox{
\includegraphics[width=0.95\columnwidth]{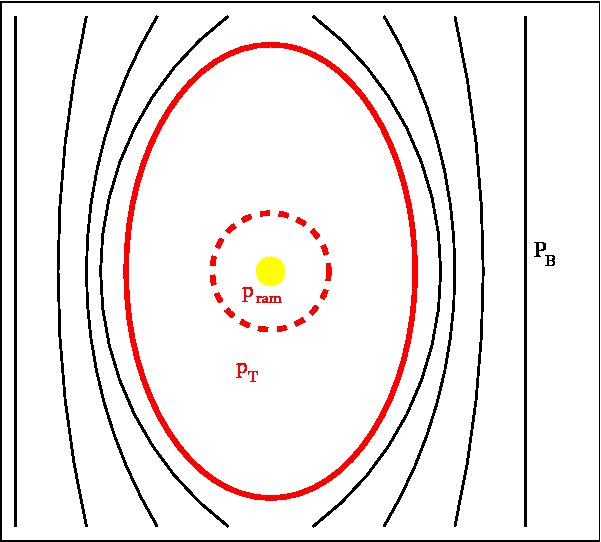}}
}
\caption{Schematic of a wind-blown bubble inside a galactic magnetic field. 
As the bubble expands it compresses the magnetic field lines (black lines), which increases the local magnetic pressure. 
The expansion of the outer shell (solid red line) stops when the thermal pressure ($\ptwind$), which drives the expansion 
is equal to the external magnetic field pressure $\pb$. This determines the location of the wind termination shock (dashed red line), 
which is located at the point where the ram pressure of the wind ($\pramwind$) is equal to $\ptwind$.
}
 \label{fig:magnetbubble}
\end{figure}

\section{The interaction between a wind-blown bubble and the interstellar magnetic field}
\label{sec-ISM}
\subsection{Hydrodynamical interaction}
The physics that govern the expansion of a wind-blown bubble are well understood.  
As the stellar wind collides with the ISM its kinetic energy is converted to thermal energy and it forms a bubble of hot, shocked gas. 
This bubble is constrained on the inside by the wind termination shock, 
where the ram-pressure of the wind ($\pramwind$) balances against the thermal pressure in the hot bubble ($\ptwind$), 
and on the outside by a shell of compressed interstellar gas that is pushed outwards 
by the thermal pressure of the bubble \citep{Weaveretal:1977}. 
(N.B. In this section, all pressures related to the ISM are denoted by an upper case $P$, 
whereas all pressures related to the wind are denoted by lower case $p$.) 
As the shell expands, its motion is determined by the balance between three different pressure sources. 
On the inside of the shell is the thermal pressure of the hot, shocked stellar wind bubble. 
On the outside the shell is confined by the thermal pressure of the ISM and the ram-pressure of its own motion into the ISM. 
Of the two confining pressures, the thermal pressure of the ISM is generally considered negligible compared to the ram-pressure. 
This is based on the assumption of a cold ISM (temperature $\sim\,100$\,K). 
The thermal pressure $\ptism$ is determined by the ISM density ($\rho_{\rm ISM}$) and the temperature ($T_{\rm ISM}$) as  
\begin{equation}
\ptism\,=\,\frac{\rho_{\rm ISM}}{\mh \mu_{\rm e}}\,k\,T_{\rm ISM},
\end{equation}
with $\mh$ the hydrogen mass, $\mu_{\rm e}$ the mean atomic weight per free 
electron 
and $k$ the Boltzmann constant. 
The ram-pressure experienced by the shell in the co-moving frame is equal to
\begin{equation}
\pramism\,=\,\rho_{\rm ISM}v_{\rm shell}^2,
\end{equation}
with $v_{\rm shell}$ the expansion speed of the shell. 
Assuming an expansion velocity of $v_{\rm shell}$ of about 10\,$\kms$ \citep[e.g.][]{vanMarleetal:2012a}, 
the ram-pressure is two orders of magnitude above the thermal pressure. 

In a warm ISM (8\,000-10\,000\,K) the situation is different, since the thermal pressure exerted by the ISM is of the same order as the ram pressure. 
Furthermore, assuming that the swept-up shell will not cool below the temperature of unshocked ISM, the shell will not be compressed. 
Under those circumstances, it is quite possible that the outward motion of the expanding bubble will become subsonic. 
If this occurs the forward shock will disappear and the shell will become an outward-moving eddy, with only a slight density increase over the ambient ISM. 
The contact contact discontinuity will remain, but rather than separating shocked wind from shocked ISM, 
its location will be determined by the pressure balance between the thermal pressure of the shocked wind and the thermal pressure of the unshocked ISM. 
A further complication arises if the warm ISM is in fact heated by the radiation from the star, which can form an HII region outside its wind-blown bubble 
as demonstrated in the numerical models of \citet{Freyeretal:2003, Freyeretal:2006, vanmarleetal:2005, vanmarleetal:2007} and \citet{ToalaArthur:2011}. 
Such an HII region expands outward into the cold ISM, even as it is being swept-up by the stellar wind, giving rise to a layered bubble consisting partially of 
shocked wind and partially of photo-ionized ISM.

\subsection{Effect of an interstellar magnetic field} 
The magnetic fields in the ambient medium can vary considerably \citep{Beck:2009}. 
In the galactic disk, field strengths of 5-10\muG\, seem to be the norm, whereas in the galactic bulge, the strength of the field 
can be more than twice as high \citep{RandKulkarni:1989,OhnoShibata:1993,Shabalaetal:2010}. 
Although these fields are not very strong in an absolute sense, they extend over large  scales (10-100\,pc).  
If the ISM contains a magnetic field, the expanding hot bubble has to overcome a constraining force comprising three separate components: 
\begin{enumerate}
 \item The thermal pressure of the ISM
 \item The ram pressure created by its own expansion into the ISM
 \item The magnetic field pressure of the ISM (for the expansion component perpendicular to the magnetic field).
\end{enumerate}
Furthermore, the nature of the outward expansion is no longer determined exclusively by the sound speed in the ISM. 
Instead, the expansion will manifest itself as a fast magnetosonic wave that travels into the ISM. 
As long as the wave is both supersonic and super-Alfv{\'e}nic, the swept-up ISM will be compressed, resembling the shells  
predicted by the purely hydrodynamical model. 
In a cold ISM the Alfv{\'e}n speed, even for a 5\muG\, magnetic field is higher than the speed of sound. 
Hence, the expansion wave will become sub-Alfv{\'e}nic, even while still supersonic. 
In terms of the shock conditions, this means that the shock makes a transition from a J-type shock to a C-type shock, because 
the fact that the Alfv{\'e}n speed exceeds the speed of sound allows the Alfv{\'e}n waves to carry information ahead of the shock. 
Once this happens, the compression will be reduced. 

Assuming that the expansion is (at least initially) super-Alfv{\'e}nic, the pressure in the compressed shell will be determined 
by a combination of thermal pressure due to shock heating and, depending on the angle of the expansion with the 
direction of the magnetic field, the magnetic pressure of the compressed magnetic field. 
If the magnetic flux that originally passed through a surface area $4\pi R^2$, with $R$ the outer edge of the expanding shell of shocked, ISM
is compressed into a shell with thickness $D$, 
the magnetic field strength in the shell ($B_{\rm shell}$) will become 
\begin{equation}
 B_{\rm shell}~=~  \frac{4\pi R^2 B_{\rm ISM}}{4\pi(R^2 -(R-D)^2)}~=~B_{\rm ISM} \frac{R^2}{2RD-D^2}, 
\end{equation}
in the direction perpendicular to the field lines. 
Because magnetic pressure scales with the field strength squared, a high compression rate ($D/R\ll1/10$), 
would indicate an increase in magnetic pressure of $\gg25$. 
This increase in magnetic pressure has to be balanced at the inner boundary of the shell by a higher thermal pressure in the shocked wind, 
which in turn requires the wind termination shock to be closer to the star. 
This also means that the expansion velocity of the hot bubble has to be \emph{reduced} (the shocked wind decompresses to drive the expansion).  
At the outer edge of the shell, the magnetic pressure can only be 
balanced by an increase in the ram pressure undergone by the expanding shell, 
which would require an \emph{increase} in expansion velocity at the forward shock. 
As a result, the swept-up shell starts to decompress. 
The amount of decompression depends on the local strength of the magnetic pressure, which in turn depends on the angle of the shell's motion with the magnetic field. 
Consequently, we can expect the contact discontinuity to follow an ovoid shape, with the major axis along the direction of the magnetic field. 
The shell will have a varying thickness, being thickest along the minor axis of the ovoid. 
Because of the decompression of the shell, the difference in magnetic field pressure between the shell and the ambient medium is reduced. 
Eventually, the shell, which by this time will have become sub-Alfv{\'e}nic, will decompress so far, that its magnetic field pressure is effectively equal to the ambient magnetic field pressure. 
At that moment we can no longer speak of a shell. 
Instead, the hot, shocked-wind bubble will be in direct pressure equilibrium with the ambient medium and the expansion perpendicular to the magnetic field stops. 
This will eventually occur for any wind-driven bubble expanding in a magnetic ISM. 
Obviously, the question is whether the star will live long enough for its bubble to reach this point. 

In order to estimate whether the circumstellar bubble will reach the point where it stops expanding perpendicular to the magnetic field 
and at what distance this will occur, we compare the two opposing forces that must reach an equilibrium: the ram pressure of the stellar wind, and the magnetic field pressure of the ISM. 
Albeit that the ram pressure of the stellar wind near the surface of a massive star is much stronger than the interstellar magnetic field,  
it is proportional to the density ($\rho$), which in a free-streaming wind decreases with $1/R^2$. 
As the stellar wind expands, it approaches the point where its ram pressure ($\pramwind$) equals the interstellar magnetic field pressure ($\pbism$). 
We can determine the distance at which this takes place:
\begin{eqnarray}
                                    \pramwind~&=&~\pbism, \\
                  \rho \vinf^2~&=&~\frac{B^2}{2\mu_0}, \\
\frac{\mdot \vinf}{4\pi \rb^2}~&=& \frac{B^2}{8\pi}, \\
                                      \rb~&=&~\frac{1}{B}\sqrt{2\,\mdot \vinf},
\end{eqnarray}
with $\vinf$ the terminal wind velocity, $B=|\Bfield|$ the magnetic field strength $\mu_0\,=\,4\pi$ the magnetic permeability of vacuum in cgs.~units, 
$\mdot$ the mass loss rate of the star and $\rb$ the radius at which the ram pressure of the stellar wind equals the magnetic field pressure in the ISM. 
Once the wind termination shock reaches this distance from the star, the ram pressure of the stellar wind (and therefore the thermal pressure of the shocked wind) 
becomes less than the interstellar magnetic field pressure. 
As a result, the bubble will stop to expand in the direction perpendicular to the magnetic field (see Fig.~\ref{fig:magnetbubble}). 
For an O-star main sequence wind ($\mdot\,=\,10^{-6}\msoy$ and $\vinf\,=\,2000\kms$) and a magnetic field strength of 5\muG, 
we find that $\rb\,\simeq\,7.3$\,pc.
This is rather large.  
However, massive stars may produce a bubble large enough to get to this point, depending on the density of the ISM. 
\citep[][found typical reverse shock radii of $\simeq\,5$\,pc for massive stars in a 10\,\#\,cm$^{-3}$ ISM at the end of their evolution.]{eldridgeetal:2006} 
Therefore, although the magnetic field pressure will influence the shape of the bubble, it will in most cases not stop the expansion. 
On the other hand, at $B\,=\,20$\muG, $\rb\,\simeq\,2.6$\,pc, placing it within the termination shock of a typical O-star or Wolf-Rayet star wind. 
   
Beside the magnetic pressure, we also have to consider the magnetic tension force. 
As long as the field  lines are parallel, the magnetic tension force is zero. 
However, when the circumstellar bubble expands, it will exert a force on the field lines, forcing them into a curved shape. 
The magnetic tension force, which is directed inward with respect to the curvature of the field lines, will try to counteract 
the expansion of the bubble. 
Therefore, the expansion of the bubble will always be reduced in the direction perpendicular to the field, 
even if the magnetic pressure is too small to overcome the ram pressure of the wind. 
   
 \begin{table}
   \label{tab:starpars}
 \centering
      \caption{
             Wind and supernova parameters, with t$_{\rm end}$ the time from the start of the simulation to the end of each evolutionary phase, 
             $\mdot$ the wind mass loss rate, $v_{\rm w}$ the wind velocity, and
             $M$ and $E$ the total mass and energy injected into the medium. 
              }
      \begin{tabular}{lccccc}
         \hline\hline
         \noalign{\smallskip}
              Phase & t$_{\rm end}$  & $\mdot$ & $v_{\rm w}$ & $M$      & $E$  \\
                    &  [\mbox{Myr}]  & [$\msoy$] & [$\kms$]    & [$\mso$] & $10^{47}$[erg] \\
         \noalign{\smallskip}
         \hline
         \noalign{\smallskip}                                                                                                  
          40$\mso$  &              &             &             &          &     \\
         \noalign{\smallskip}
         \hline
         \noalign{\smallskip}                                                                                                  
          MS       & 4.3  & 1.0$\times10^{-6}$  & 2\,000  & 4.3  & 1\,710    \\
          RSG      & 4.5  & 5.0$\times10^{-5}$  & 15      & 10.0 & 0.224     \\
          WR       & 4.8  & 1.0$\times10^{-5}$  & 2\,000  & 2.7  & 1\,190    \\
          SN       &      &                     &         & 10.0 & 10\,000   \\
         \noalign{\smallskip}
         \hline
      \end{tabular}
   \end{table}

\section{Method}
\label{sec-method}
\subsection{Computational method}
We use the \emph{MPI-AMRVAC} hydrodynamics code \citep{Keppensetal:2012}, 
which solves the conservation equations for mass, 
\begin{equation}
\frac{\partial \rho}{\partial t} ~+~ \nabla \cdot (\rho \vel) ~=~ 0,
\label{eq:mass}
\end{equation}
momentum:  
\begin{equation}
\rho \biggl(\frac{\partial \vel}{\partial t}+\vel\cdot\nabla\vel\biggr)~+~\nabla p_{\rm tot} -\frac{1}{\mu_0}(\nabla\times\Bfield)\times\Bfield~=~0,
\label{eq:momentum}
\end{equation}
and energy: 
\begin{equation}
\frac{\partial e}{\partial t}+\nabla \cdot (e {\vel} ) ~+~ \nabla \cdot (p_{\rm tot}\vel) ~=~ -\biggl(\frac{\rho}{m_h}\biggr)^2 \Lambda(T), 
\label{eq:energy}
\end{equation}
with $\vel$ the velocity vector, $p_{\rm tot}$ the sum of the thermal pressure and the magnetic pressure, and 
$m_h$ the hydrogen mass. 
The energy density $e$ is defined as the sum of thermal, kinetic and magnetic energy density,
\begin{equation}
e~=~\frac{p}{\gamma-1}~+~\frac{\rho v^2}{2}~+~\frac{B^2}{2\mu_0}, 
\end{equation}
with $\gamma=5/3$ the adiabatic index and $v=|\vel|$. 
We assume ideal MHD, so there are no local source terms for the magnetic field. 
Consequently,
\begin{equation}
\frac{\partial \Bfield}{\partial t}~-~\nabla\times({\vel \times \Bfield})~=~0. 
\end{equation}
We solve these equations using the total variation diminishing, Lax-Friedrich {\tt TVDLF} method combined with a koren flux limiter \citep{Kuzmin:2006}, 
except for the two highest levels of refinement, 
where we apply the more diffusive minmod flux limiter \citep{Roe:1986} to prevent the growth of numerical instabilities at the polar axis.

The divergence of the magnetic field is kept at zero with the method described by \citet{Powelletal:1999}.
Equation~\ref{eq:energy} includes the effect of radiative cooling, which depends on local density, 
as well as a temperature dependent cooling curve for solar-metallicity gas, $\Lambda(T)$, 
obtained from \citet{Schureetal:2009}. 
This curve, generated with the \emph{SPEX} code,
covers a temperature range of $10^4$-$10^8$\,K. 
For higher temperatures, we assume that the cooling scales with $\sqrt{T}$, because the Bremsstrahlung dominates the cooling. 
For temperatures below $T\,=\,10^4$\,K, we copy the approach of \citet{Schureetal:2009}, by using the cooling function 
from \citet{DalgarnoMcCray:1972}, adjusted for a pre-set ionization fraction, which we choose to be $10^{-3}$. 
We set a minimum temperature of 100\,K throughout the simulation to prevent numerical problems that can arise because of 
extreme compression of radiatively cooling gas. 

The simulations are set up in a manner similar to the one used in \citet{vanMarleetal:2012a}, 
with a 2D cylindrical grid, which is symmetric around the z-axis. 
We start with a basic resolution of 0.35\,pc per grid cell and allow 3 additional levels of refinement, each doubling the resolution. 
This gives us an effective resolution of 0.044\,pc. 
This grid is filled with ambient medium according to the parameters specified in Sect.~\ref{sec-ismpars}. 
The stellar wind is introduced in the grid by filling a small half-sphere (0.25\,pc in radius) on the z-axis with wind material, 
with the wind parameters changing over time according to the evolutionary phase of the star (Table~1). 
Around this sphere we enforce an additional level of grid refinement (0.022\,pc per cell) 
to reduce numerical errors, a possible occurrence, when a spherical expansion is simulated on a rectangular grid. 
The stellar wind parameters are the same as those used for the 40$\mso$ star in our earlier paper \citep{vanMarleetal:2012a}. 
The mass loss rates and evolution time are based on the model used in \citet{vanmarleetal:2005}. 
Wind velocities reflect those typically found in observations \citep[][and references therein]{Lamerscassinelli:1999}. 
The mass-loss rate in the final phase (the Wolf-Rayet phase) has been reduced 
to reflect the lower values obtained by \citet{Vinkdekoter:2005} when taking into account the effects of clumping. 

The supernova is introduced in a similar manner: during the first time step of the supernova simulation, the same sphere is filled with 
hot gas with a total mass of $10\mso$ and a total thermal energy of $10^{51}$erg. 
The supernova mass reflects the final mass of the star as well as the values calculated by, e.g., \citet{EldridgeTout:2004}. 
For simplicity, we use a supernova energy of $10^{51}$erg, 
which is typical of core-collapse supernovae \citep[e.g.][and references therein]{vanMarleetal:2010}. 
During the supernova phase we allow the maximum level of resolution (0.022\,pc per cell) 
throughout the entire domain to make sure we resolve the high velocity shock front.

Throughout the simulations, the interstellar magnetic field $B_{\rm ISM}$ is set parallel with the z-axis and fixed at the outer boundaries, 
using the same method as in \citet{vanMarleetal:2014,vanMarleetal:2014b}.
We do not take into consideration a stellar magnetic field. 
Simulations by \citet[][]{uddoulaowocki:2002} and \citet{uddoulaetal:2008,uddoulaetal:2009} show that the magnetic fields of massive stars, 
important when close to the stellar surface, 
are typically dominated by the ram pressure of the wind at large distances. 

\subsection{ISM parameters}
\label{sec-ismpars}
We run four simulations varying only the magnetic field strength in the ISM:
\begin{itemize}
 \item Simulation~A, the ``basic'' model. A 40$\mso$ star in an ambient medium of $10^{-22.5}\gcm\,\simeq\,20\,$\#\,cm$^{-3}$ without a magnetic field. 
The ambient medium temperature is set to 100\,K.
This model serves as a reference against which we measure the other models. 
 \item Simulation~B, similar to simulation~A, but with an interstellar magnetic field of 5\muG, 
 which is representative of the weak magnetic fields in the galactic disk. 
 \item Simulation~C, similar to simulation~B, but with an interstellar magnetic field strength of 10\muG, 
 representative of the strong magnetic fields in the galactic disk and the weak fields in the galactic bulge. 
 \item Simulation~D, similar to simulations~B and C, but with an interstellar magnetic field strength of 20\muG, 
 representative of the strong magnetic fields in the galactic bulge. 
\end{itemize}
These allow us to explore both the qualitative and quantitative effects of the influence of the interstellar magnetic field on the morphology of the circumstellar bubble, 
 and, additonally, we run two simulations for an ISM with a temperature of 10\,000\,K to show how the effect of the interstellar magnetic field 
 combines with other forces acting on the bubble:
\begin{itemize}
\item  Simulation~E, similar to simulation~B but with an interstellar density of $10^{-24.5}\gcm$ and an ISM temperature of 10\,000\,K as can be expected in warm, low density ISM.
\item  Simulation~F, similar to simulation~B, but an ISM temperature of 10\,000\,K, reminiscent of the conditions in an H$_{\rm II}$ region.
\end{itemize}

\begin{figure*}
\FIG{
 \centering
\mbox{
\subfigure
{\includegraphics[width=0.33\textwidth]{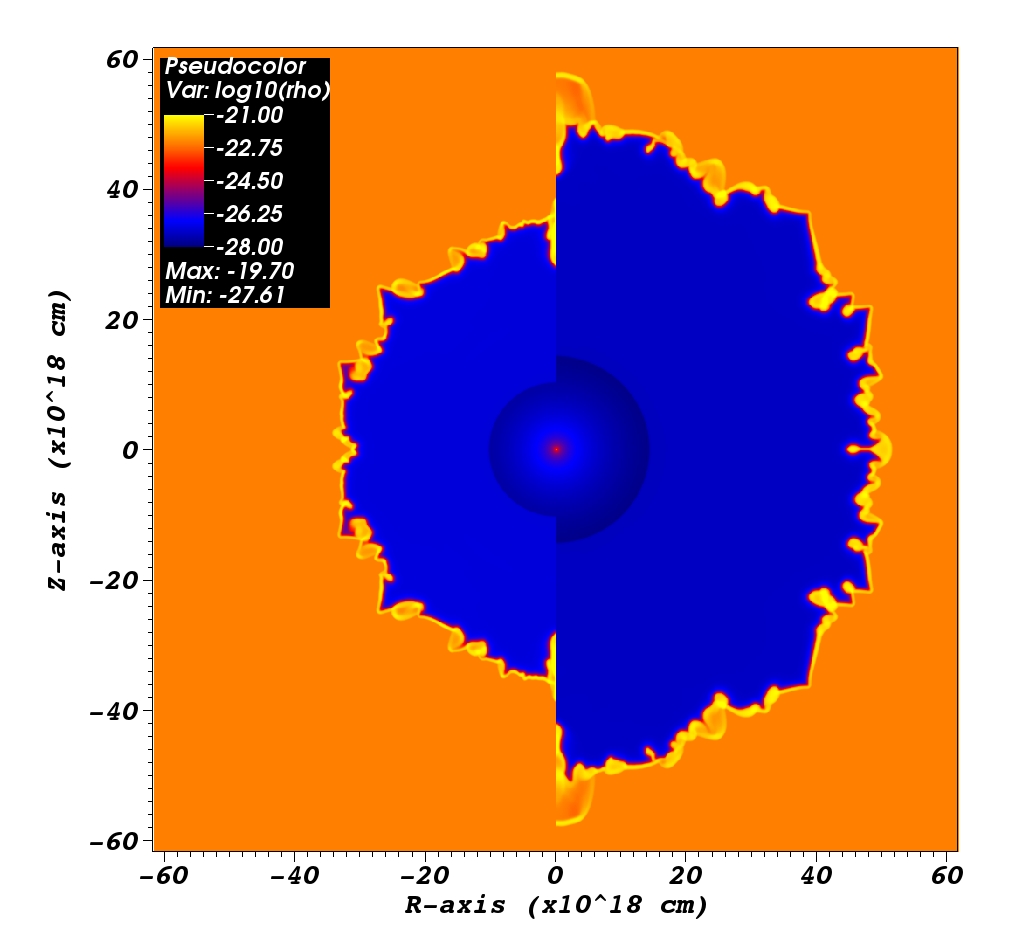}}
\subfigure
{\includegraphics[width=0.33\textwidth]{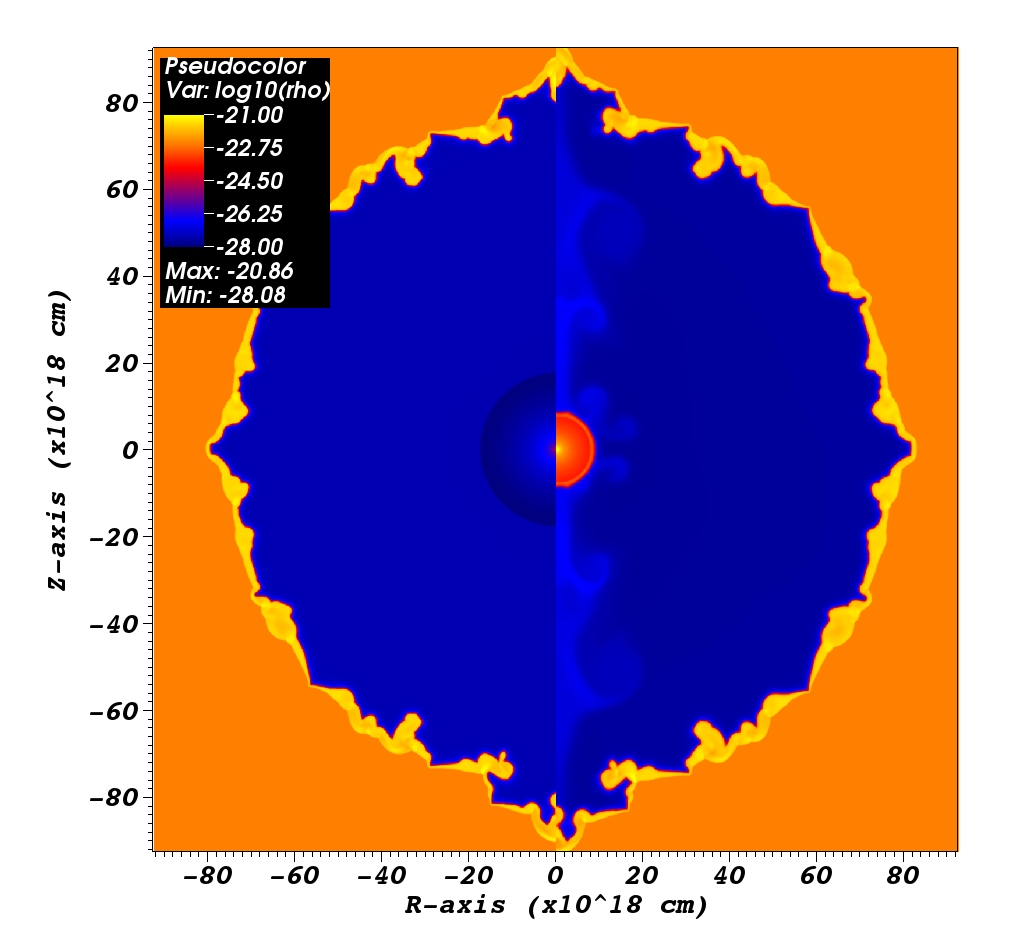}}
\subfigure
{\includegraphics[width=0.33\textwidth]{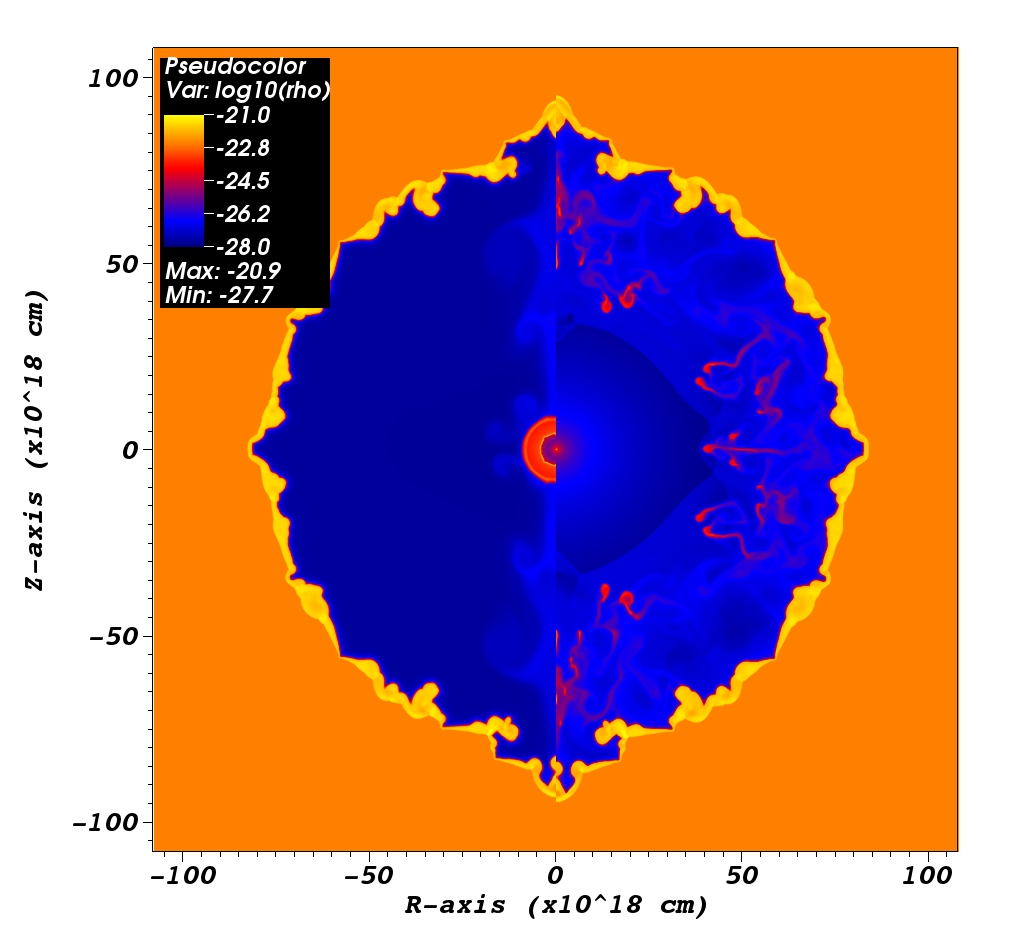}}}
}
\caption{Logarithm of the density [cgs.] for the circumstellar medium in the absence of an interstellar magnetic field. 
The left panel shows the circumstellar bubble on a 40$\times$40\,pc frame after 1 (left side) and 2 (right side) Myr have past since the start of the main sequence. 
The bubble is spherically symmetric, except for local thin-shell and Rayleigh-Taylor instabilities.  
In the centre (frame size 60$\times$60\,pc), the star leaves the main sequence after 4.3\,Myr (left side) and becomes a red supergiant, which lasts  
until 4.5\,Myr after the start of the simulation (right side), forming a new shell at the wind termination shock. 
The right panel (frame size 70$\times$70\,pc) shows the transition to the Wolf-Rayet stage, which forms a new shell that sweeps up the red supergiant wind (left panel, 
10\,000\,yrs into the Wolf-Rayet stage) and collides with the red supergiant shell. 
The fragments of this collision move out into the old main sequence bubble (right side, 100\,000 years into the Wolf-Rayet fase).
}
\label{fig:wsnB0_fig1}
\end{figure*}

\begin{figure*}
\FIG{
 \centering
\mbox{
\subfigure
{\includegraphics[width=0.33\textwidth]{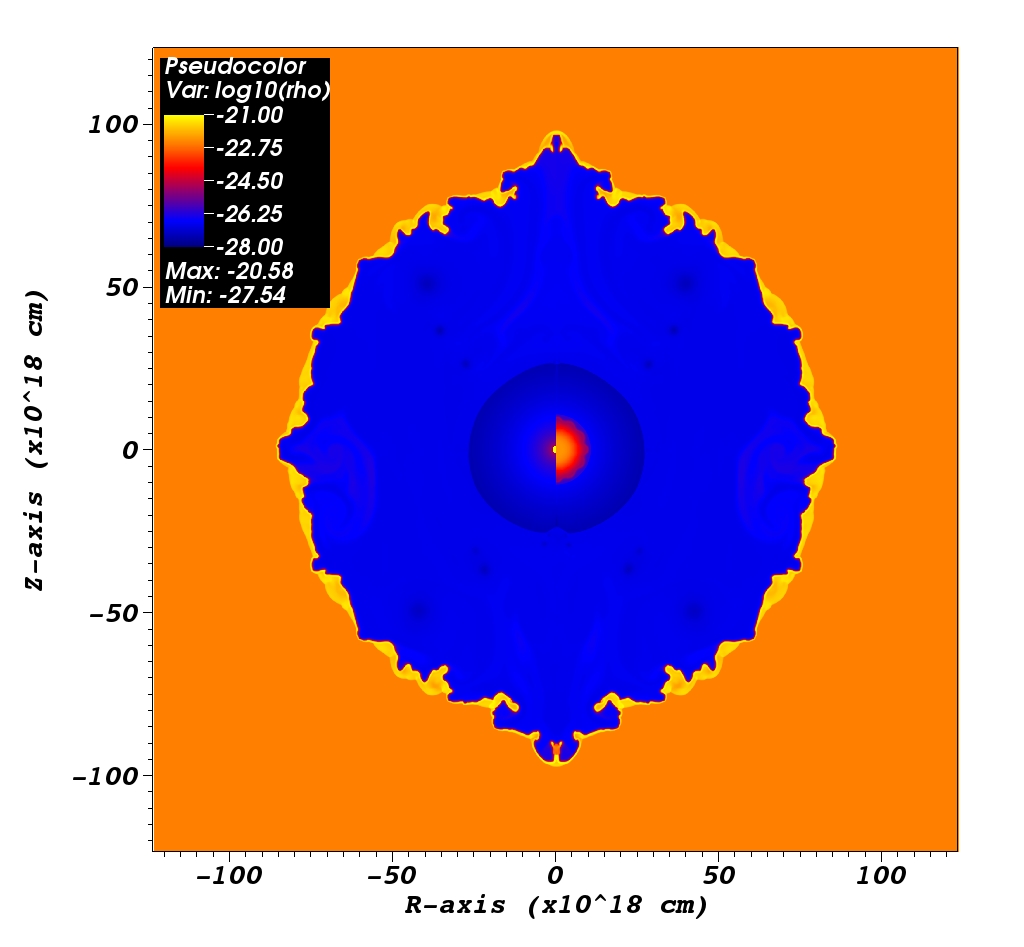}}
\subfigure
{\includegraphics[width=0.33\textwidth]{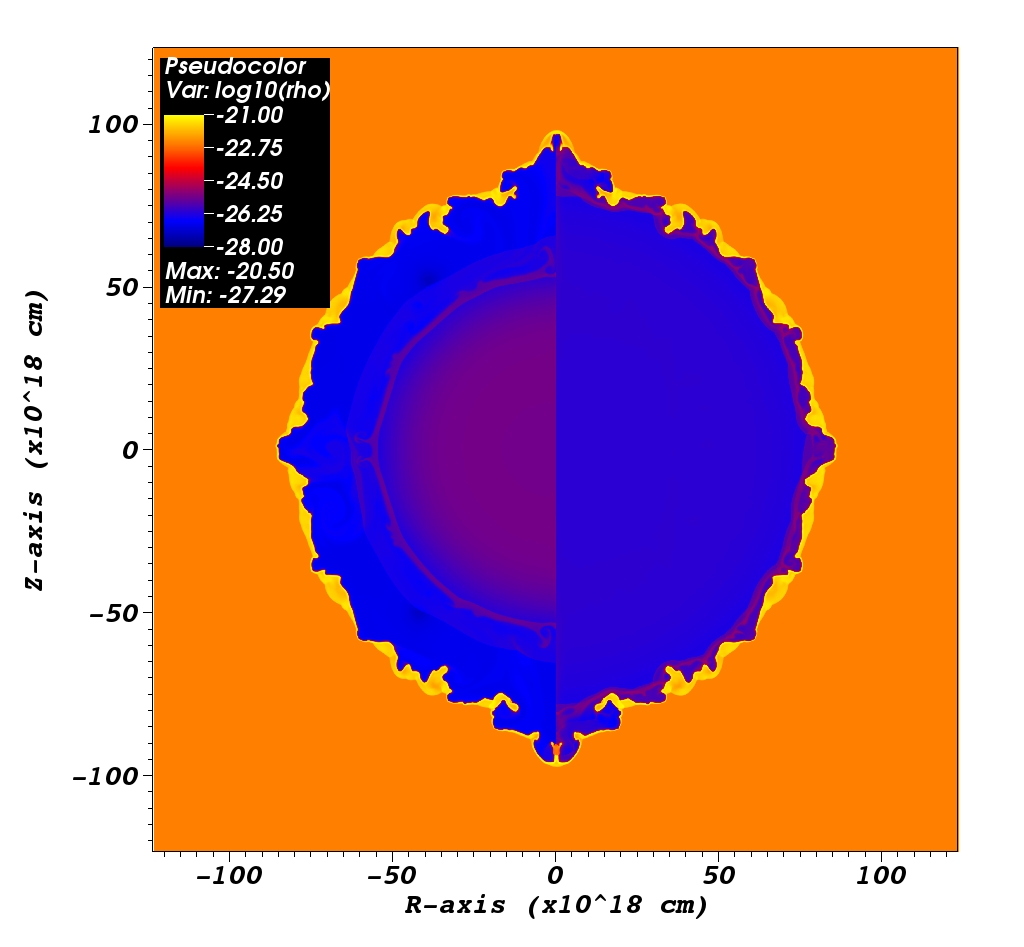}}
\subfigure
{\includegraphics[width=0.33\textwidth]{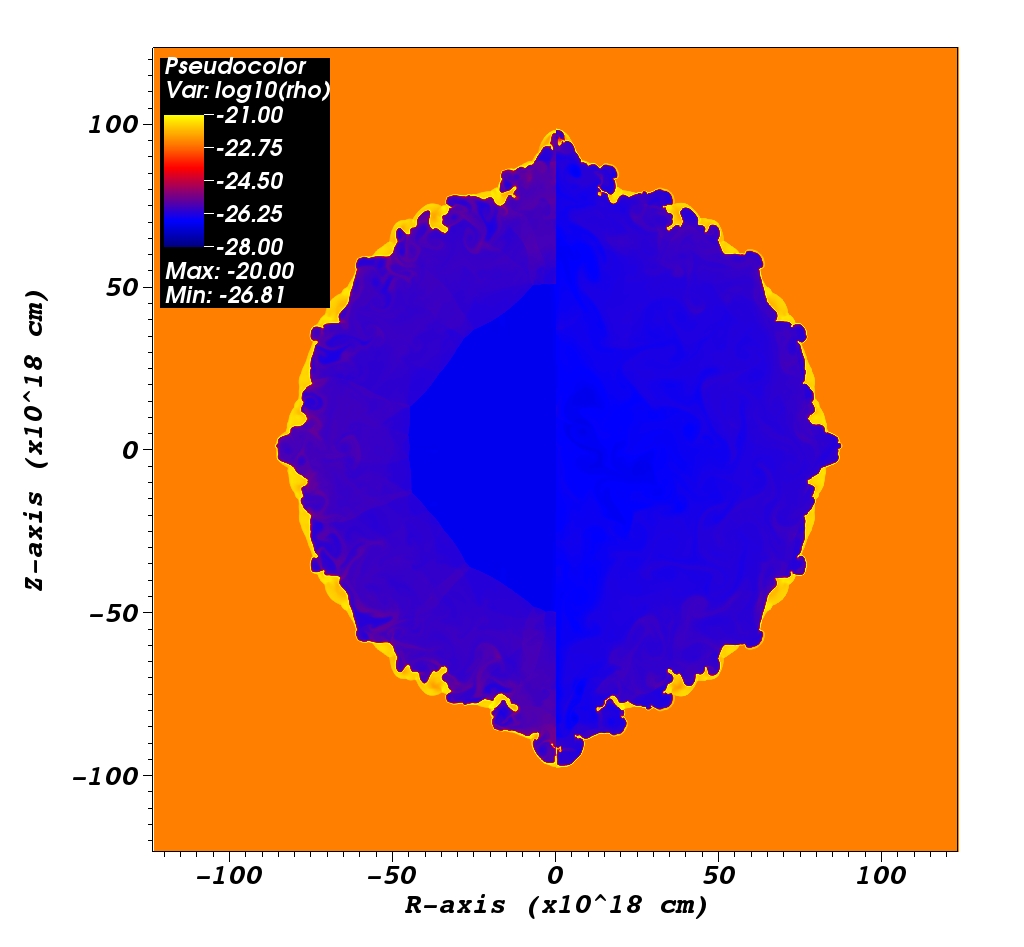}}}
}
\caption{Similar to Fig.~\ref{fig:wsnB0_fig1}, showing the expansion of the supernova remnant. 
All frames have the same size of $80\times80$\,pc. 
At the end of the Wolf-Rayet phase (left side of left panel) the density in the hot bubble has become smooth once again. 
The supernova starts as a high pressure blast wave at the centre of the bubble and quickly expands.  
After 400 years it has swept up nearly half of the free-streaming Wolf-Rayet wind (right side of left panel). 
4000 years  after the explosion (left side of centre panel), the supernova remnant has swept up most of the hot bubble. 
After 7200 years (right side, centre panel) it collides with the outer shell. 
Afterwards, the supernova remnant flows back toward the centre of the bubble (12\,000 years after the explosion, left side of right panel). 
Eventually, its motion will slow down to subsonic level and the pressure in the bubble will start to equalize (right side of right panel, 40\,000 years after the explosion). 
}
\label{fig:wsnB0_fig2}
\end{figure*}

\section{Bubbles in a cold ISM}
\label{sec-results}
\subsection{Expansion in the absence of a magnetic field}
Simulation~A shows our basic model, which was also used in \citet{vanMarleetal:2012a}. 
The evolution of the circumstellar bubble follows the expected path.
The main sequence wind creates a large, spherical bubble of hot, shocked wind material, which expands into the ambient medium, 
sweeping up a shell, following the pattern predicted by the analytical model \citep{Weaveretal:1977}. 
This is shown in the left panel of Fig.~\ref{fig:wsnB0_fig1}. 
The shell is subject to instabilities, primarily of the linear thin shell type \citep{Vishniac:1983}, in the earlier stages. 
These instabilities result from the different forces acting on the shell: isotropic thermal pressure on the inside and ram-pressure from the outside. 
Rayleigh-Taylor instabilities, resulting from the density difference at the contact discontinuity between the shocked wind and the swept-up shell develop more slowly, 
but begin to dominate in the later stages of the main sequence (right side of the left panel and left side of the centre panel of Fig.~\ref{fig:wsnB0_fig1}). 

As the star makes the transition to the RSG phase, the wind parameters change, leading to a change in the inner part of the CSM. 
The high density RSG wind forms a shell at the wind termination shock (right side of the centre panel of Fig.~\ref{fig:wsnB0_fig1}).
Eventually, the star becomes a Wolf-Rayet star, with a fast, powerful wind that sweeps up the RSG wind material in a shell 
(left side of the right panel of Fig.~\ref{fig:wsnB0_fig1}). 
This Wolf-Rayet wind driven shell collides with the RSG wind shell and both shells are fragmented. 
The individual fragments move out into the shocked main sequence wind bubble, where they dissipate or merge with the outer shell 
(right side of the right panel of Fig.~\ref{fig:wsnB0_fig1}). 

This sequence of events was shown previously by \citet[e.g.][]{GarciaSeguraetal:1996b, vanmarleetal:2005, Freyeretal:2006, 
Dwarkadas:2005} and \citet{ToalaArthur:2011} and can also be seen in our animations (Sect.~\ref{sec-animations}). 

\begin{figure}
\FIG{
 \centering
\mbox{
\includegraphics[width=\columnwidth]{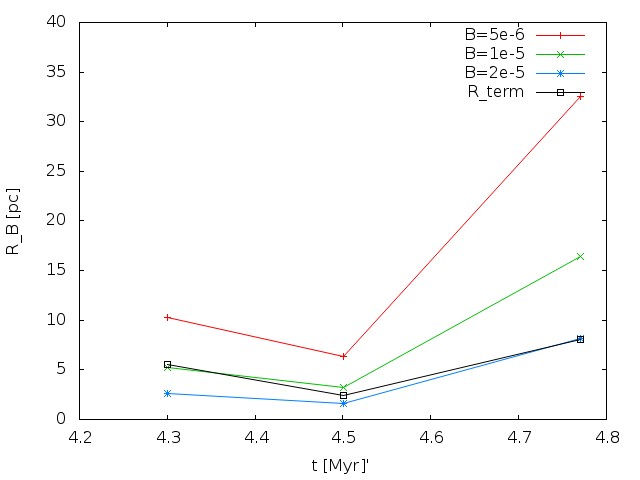}}
}
\caption{Values for the wind termination shock obtained from simulation~A (black) at the end of each evolutionary phase, 
compared to $\rb$ for the three magnetic field strengths. 
At 5\muG\, (red), the magnetic field can never stop the expansion of the circumstellar bubble perpendicular to the field  
and the 10\muG\, (green) field will do so, but only at the very end of the main sequence. 
The 20\muG\, (blue) field will stop expansion throughout most of the stellar evolution. 
}
 \label{fig:rb_plot}
\end{figure}

Using the results from this simulation we can make a rough prediction as to what extent the magnetic fields in the subsequent simulations 
will inhibit the expansion of the circumstellar bubble. 
Figure ~\ref{fig:rb_plot} shows the wind termination shock at the end of each evolutionary phase, compared 
to the $\rb$ values for the stellar wind parameters and various magnetic field strengths. 
As long as the termination shock radius is smaller than $\rb$ the wind driven bubble will be able to expand perpendicularly to the magnetic field, 
albeit at a lower speed owing to the magnetic tension force and the increased pressure from the ISM.  
This is what we can expect for simulations~B and C. 
For simulation~D, which has an $\rb$ that is lower than the termination shock for most of the simulation, 
expansion perpendicular to the field will stop completely. 

Once the star reaches the end of its evolution, it explodes as a core-collapse supernova (left panel of Fig.~\ref{fig:wsnB0_fig2}). 
This causes a sudden injection of mass and energy at the centre of the circumstellar bubble, which by now has a radius of approximately 30\,pc. 
The supernova remnant expands quickly into the low density medium until it reaches the inner edge of the swept-up ISM shell (centre panel of Fig.~\ref{fig:wsnB0_fig2}).  
There it is stopped by the high mass ($\sim\,40\,000\,\mso$) shell. 
Failing to break out, the supernova remnant recoils and starts an inward motion (left side of right panel of Fig.~\ref{fig:wsnB0_fig2}), 
which continues until it reaches the centre of the bubble. 
From this point the movement slows down until, finally, the density and pressure inside the bubble start to even out (right side of right panel of Fig.~\ref{fig:wsnB0_fig2}).

\begin{figure*}
\FIG{
 \centering
\mbox{
\subfigure
{\includegraphics[width=0.33\textwidth]{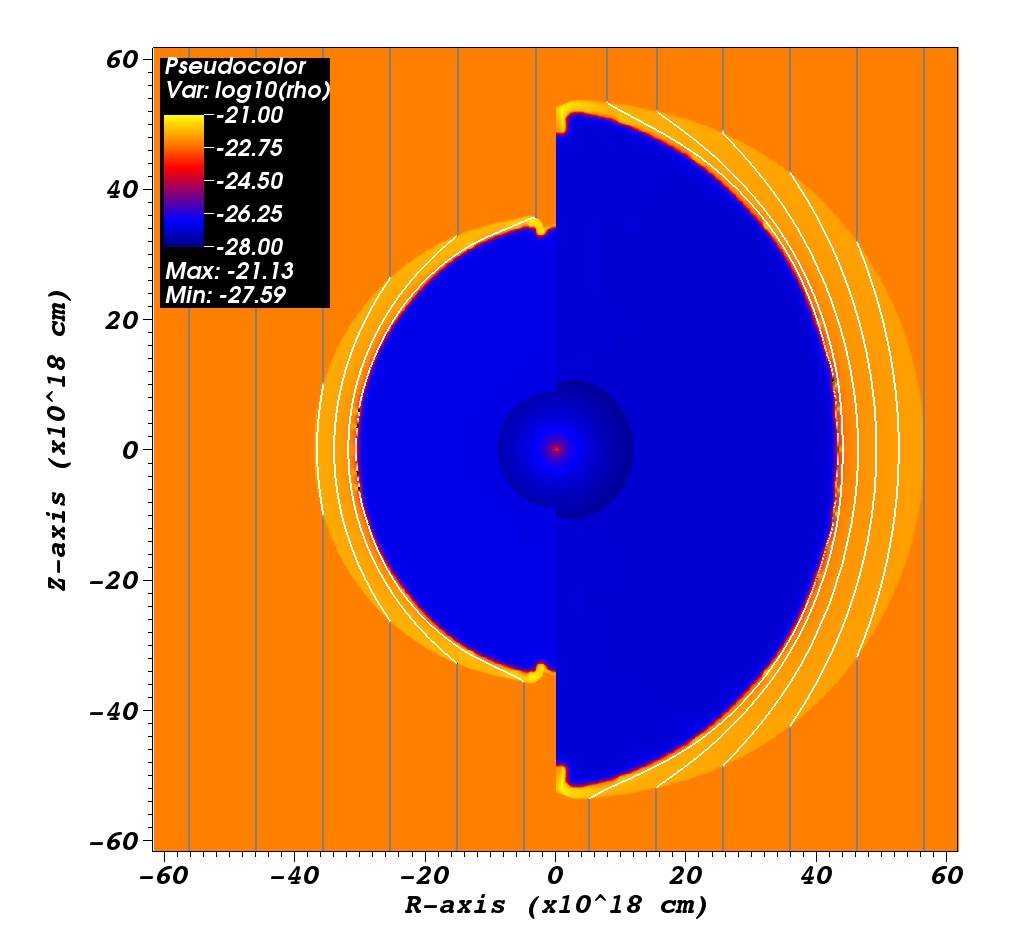}}
\subfigure
{\includegraphics[width=0.33\textwidth]{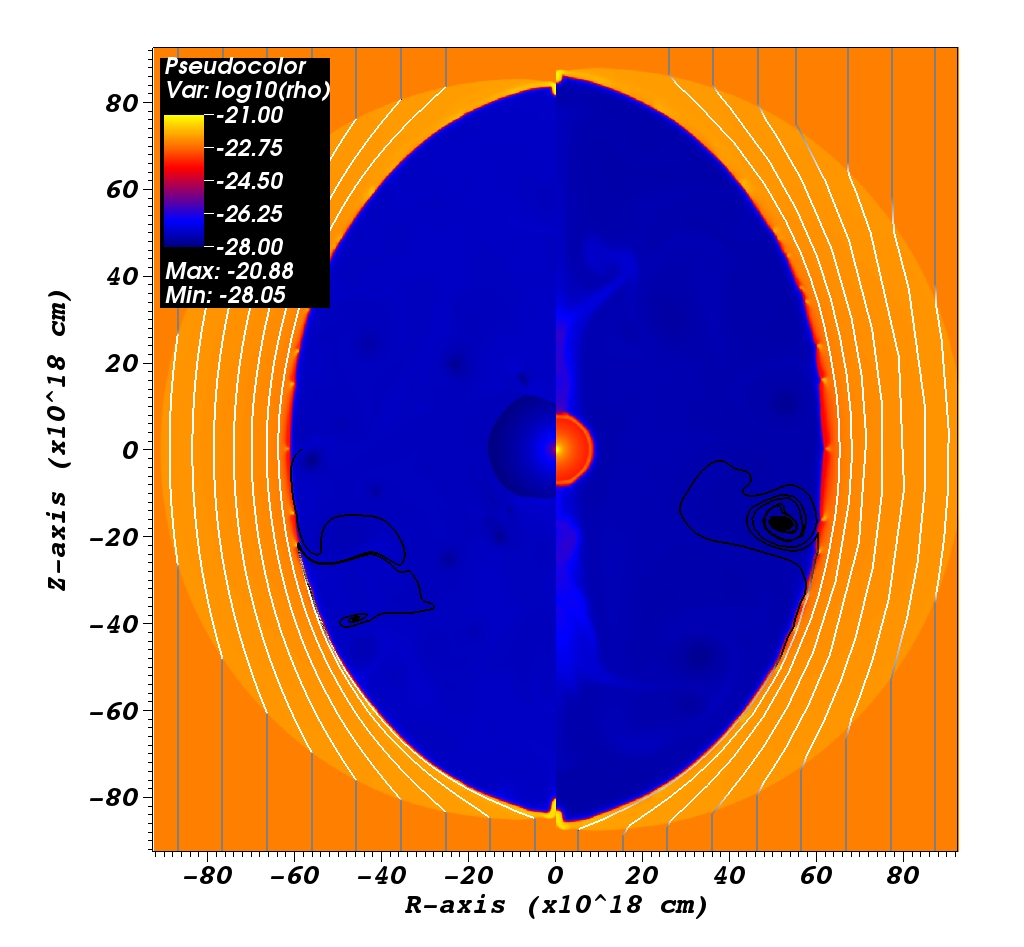}}
\subfigure
{\includegraphics[width=0.33\textwidth]{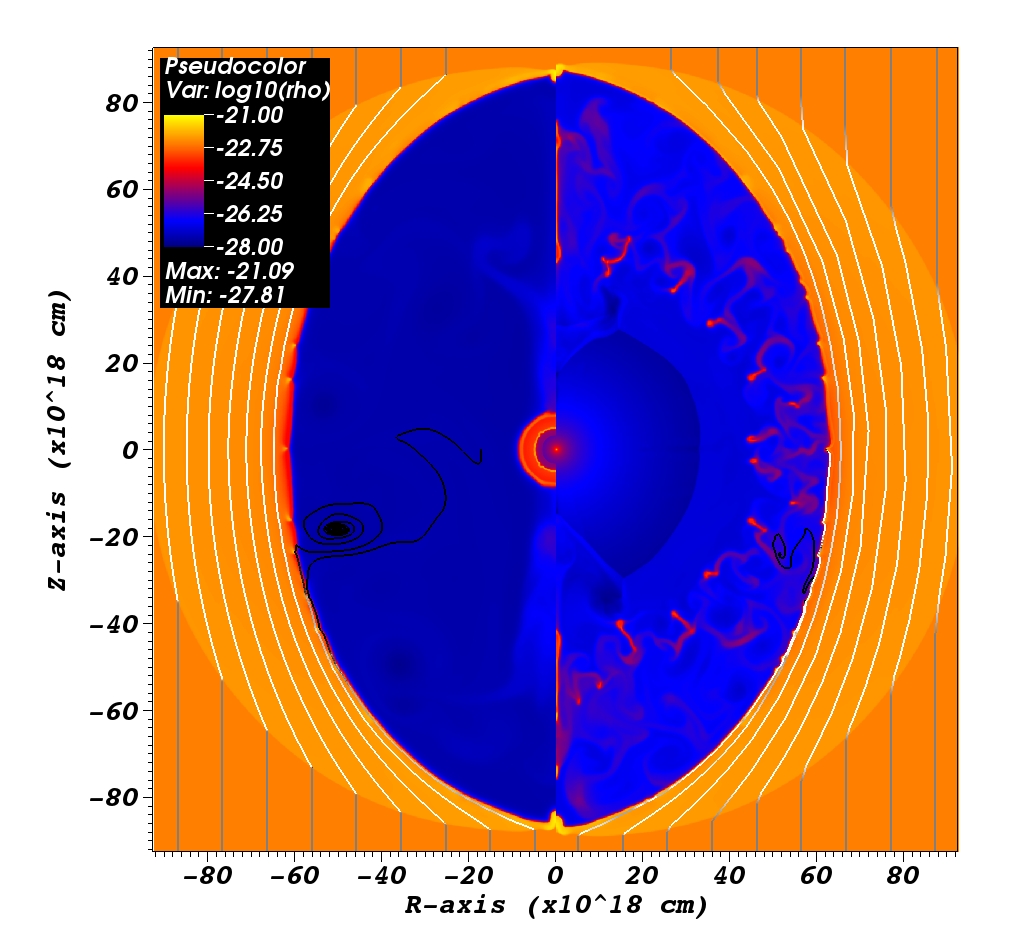}}}
}
\caption{Similar to Fig.~\ref{fig:wsnB0_fig1}, but for Simulation~B with an external magnetic field of 5\muG, showing the progression of the bubble evolution 
at 1, 2, 4.3, 4.5, 4.51 and 4.6~Myr. 
Unlike the non-magnetic simulation, the swept-up shell of shocked ISM is relatively thick, 
which prevents the formation of thin-shell instabilities. 
The overall-shape of the bubble is ellipsoid owing to the magnetic pressure which slows down expansion perpendicular to the field, 
but the evolution of temporary structures inside the bubble is similar to the non-magnetic case. 
A faint sign of instability can be seen along the contact discontinuity, but the instabilities do not grow over time. 
}
\label{fig:wsnB5_fig1}
\end{figure*}

\begin{figure*}
\FIG{
 \centering
\mbox{
\subfigure
{\includegraphics[width=0.33\textwidth]{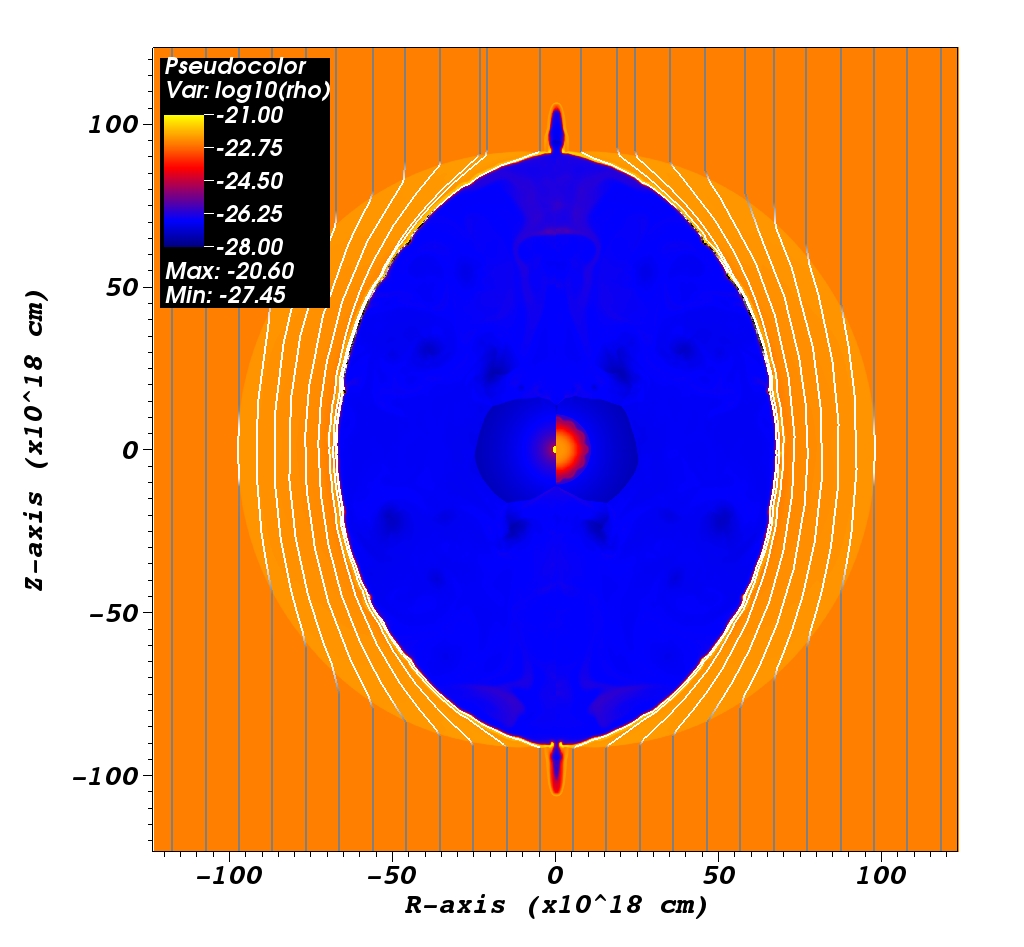}}
\subfigure
{\includegraphics[width=0.33\textwidth]{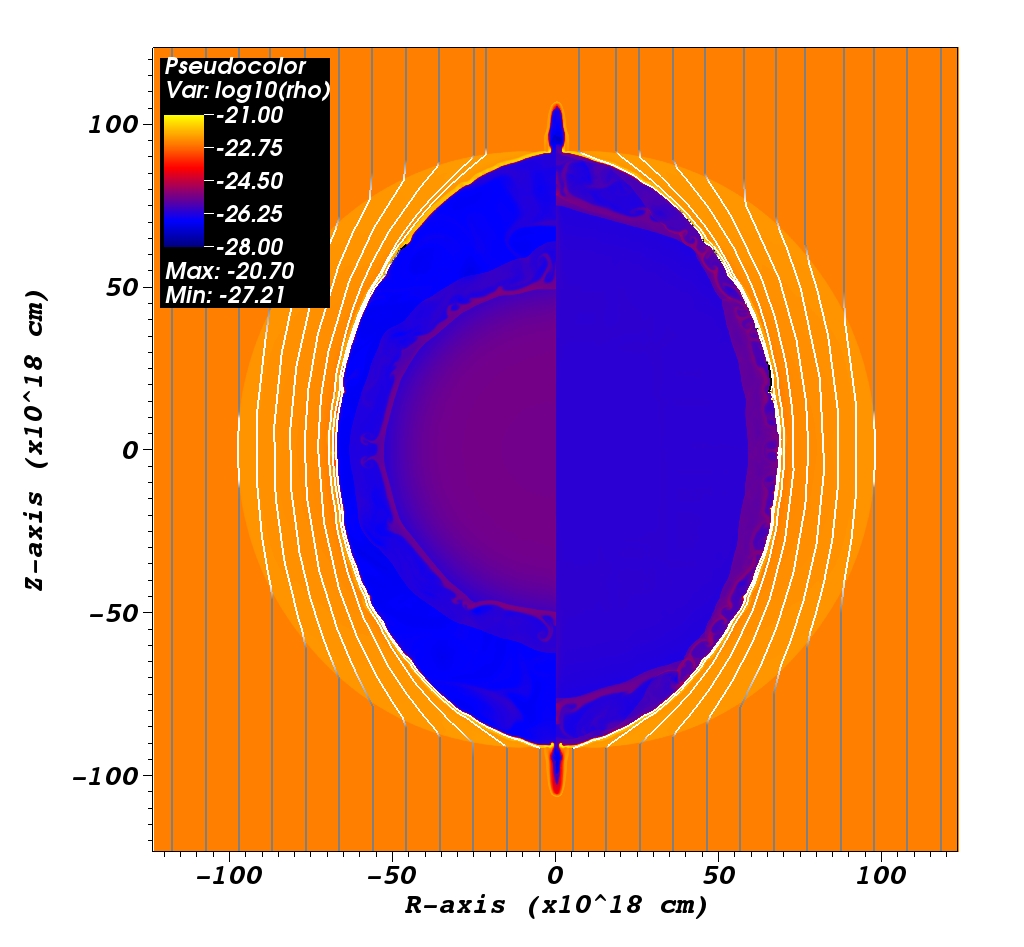}}
\subfigure
{\includegraphics[width=0.33\textwidth]{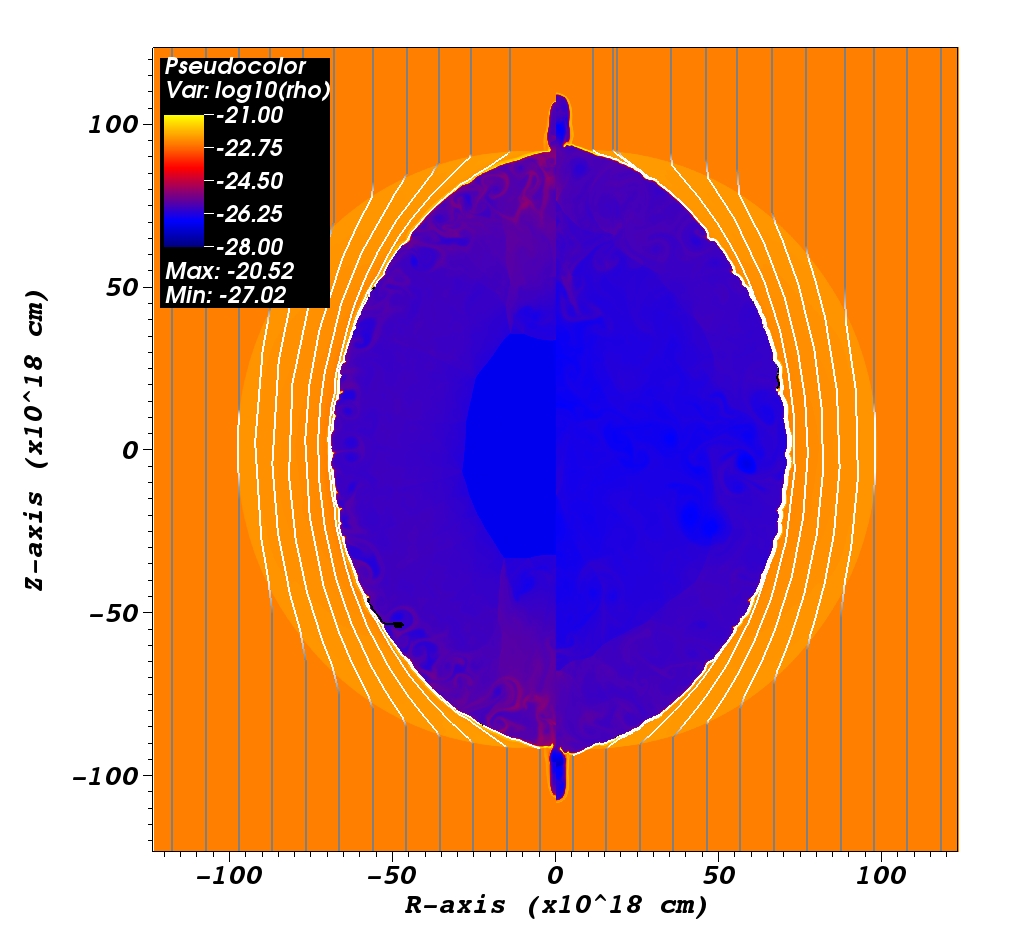}}}
}
\caption{Similar to Fig.~\ref{fig:wsnB0_fig2}, but for Simulation~B with an external magnetic field of 5\muG. 
Initially, the supernova expansion is nearly identical to the non-magnetic case, but 
since the supernova expansion is spherical and the outer boundary of the bubble is ellipsoid, 
the supernova will first collide with outer shell at the ``waist'' of the bubble. 
}
\label{fig:wsnB5_fig2}
\end{figure*}

\subsection{Low interstellar magnetic field: 5\muG}
Simulation~B shows the evolution of the circumstellar bubble in a weak (5\muG) external magnetic field, 
representative of the magnetic field in the spiral arms of the galactic disk on 10-100\,pc scales 
\citep{RandKulkarni:1989,OhnoShibata:1993,Shabalaetal:2010,Greenetal:2012} and the lower values obtained for magnetic fields in nearby galaxies \citep{Beck:2009,Fletcheretal:2011}. 
During the main sequence phase, the bubble expands outward, 
creating a non-magnetic hole in the field by pushing the field lines, 
embedded in the ISM, outwards (left and centre panels of Fig~\ref{fig:wsnB5_fig1}). 
The magnetic tension force, which is directed inward with respect to the curvature, 
counteracts the displacement of the field lines, reducing the expansion perpendicular to the field lines. 
At the same time the field lines are pressed together in the shocked shell, increasing the magnetic pressure, 
but its strength is insufficient to stop the expansion. 
This results in an ellipsoid circumstellar bubble. 

Owing to the asymmetrical shape of the bubble, the remnants of the WR-RSG shell collision will first encounter the outer shell at the point where it has expanded least 
(the direction perpendicular to the external field). 
This causes the shell fragments to be pushed aside along the z-direction, causing a mirror symmetry, but most of the turbulence disappears 
before the star reaches the end of its evolution (right panel of Fig.~\ref{fig:wsnB5_fig1} and left side of left panel of Fig.~\ref{fig:wsnB5_fig2}. 
At the end of the stellar evolution (left side of left panel of Fig.~\ref{fig:wsnB5_fig2})), 
the shocked wind bubble is elliptical, with an asymmetry of approximately 1.3:1, 
because the magnetic field pressure limits the expansion in the direction perpendicular to the field (r-direction). 
Along the axis parallel with the magnetic field, the bubble has a radius of approximately, 30\,pc, comparable with the radius of the bubble in simulation~A, 
whereas the radius along the shorter axis is approximately 23\,pc. 

The magnetic field lines that are swept up in the outer shell resist being pressed together, consequently the 
compression in the shell is lower than in the model without a magnetic field. 
This results in a shell with a variable thickness: widest where the shell moves perpendicular to the field and thinnest where it moves parallel to the field,  
resulting in the outer edge of the shell being nearly spherical, despite the elliptical shape of the contact discontinuity. 
The shell lacks instabilities, because the magnetic pressure makes it too thick for thin-shell instabilities (Fig~\ref{fig:wsnB5_fig1}). 
Rayleigh-Taylor instabilities could still develop, but they would have to move perpendicular to the magnetic field lines. 
Such instabilities can be seen faintly along the contact discontinuity (best observed in centre panel of Fig.~\ref{fig:wsnB5_fig1}), but the 
instabilities do not grow over time.

Eventually, the star explodes as a supernova. 
The evolution of the supernova remnant (Fig.~\ref{fig:wsnB5_fig2}) closely resembles that of the non-magnetic models,  
since there is no direct contact between the magnetic field and the supernova remnant, which is contained by the shocked wind bubble. 
Like the expanding WR shell, the supernova first encounters the outer shell at the ``waist'' of the bubble (right side of centre panel in Fig.~\ref{fig:wsnB5_fig2}), 
causing a mirror symmetric structure inside the hot bubble. 
Over time this structure tends to disappear until, eventually, the density inside the bubble becomes nearly constant (right panel of Fig.~\ref{fig:wsnB5_fig2}).

\begin{figure*}
\FIG{
 \centering
\mbox{
\subfigure
{\includegraphics[width=0.33\textwidth]{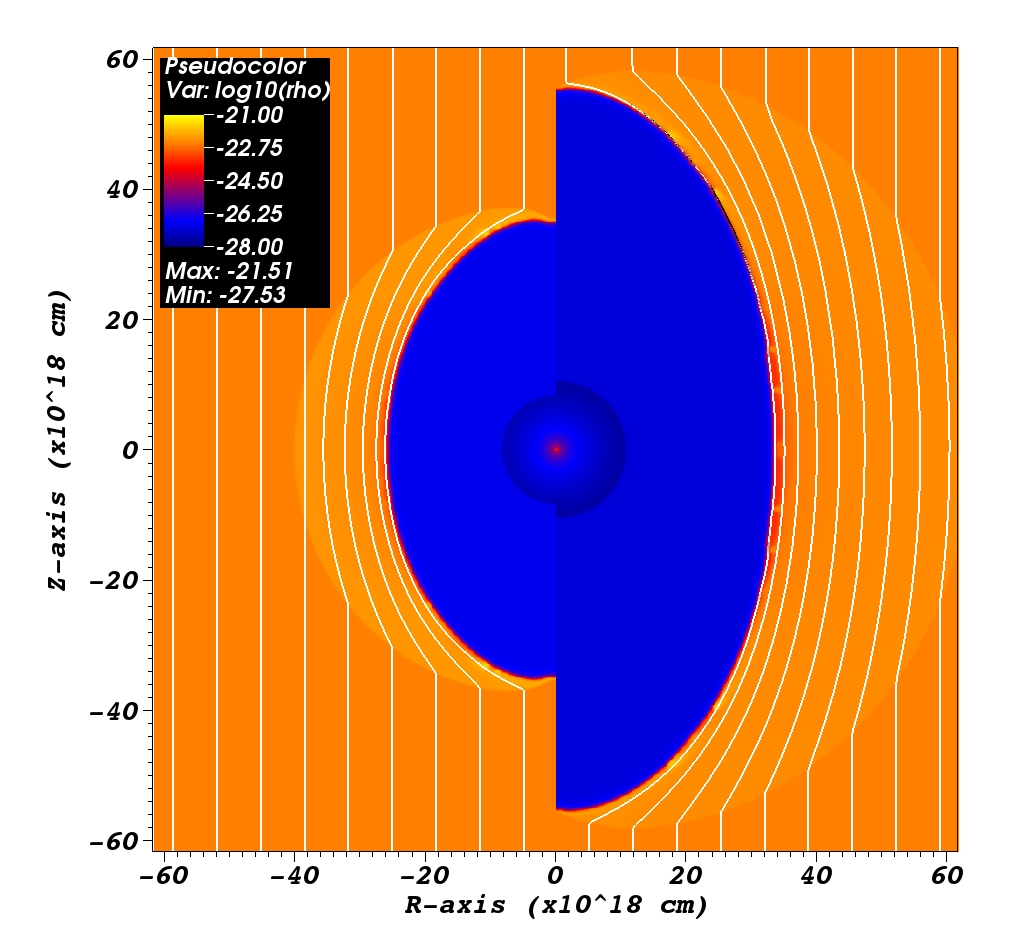}}
\subfigure
{\includegraphics[width=0.33\textwidth]{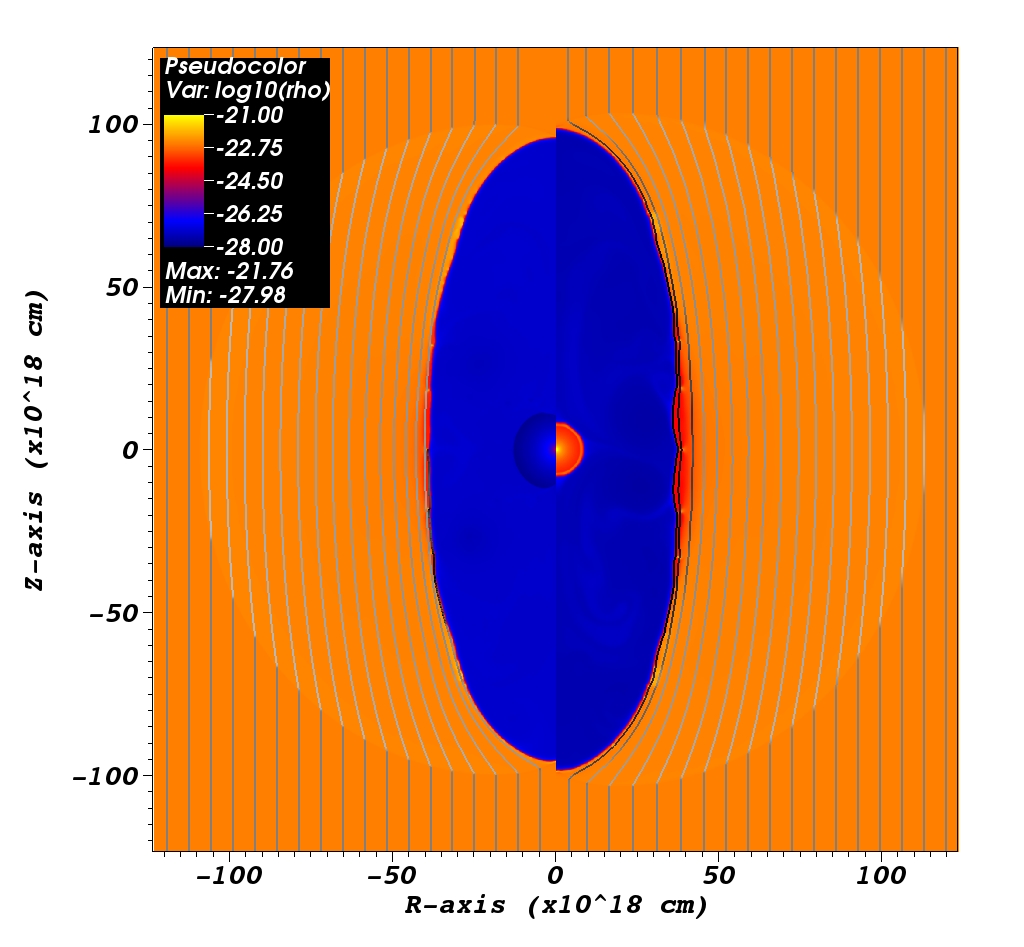}}
\subfigure
{\includegraphics[width=0.33\textwidth]{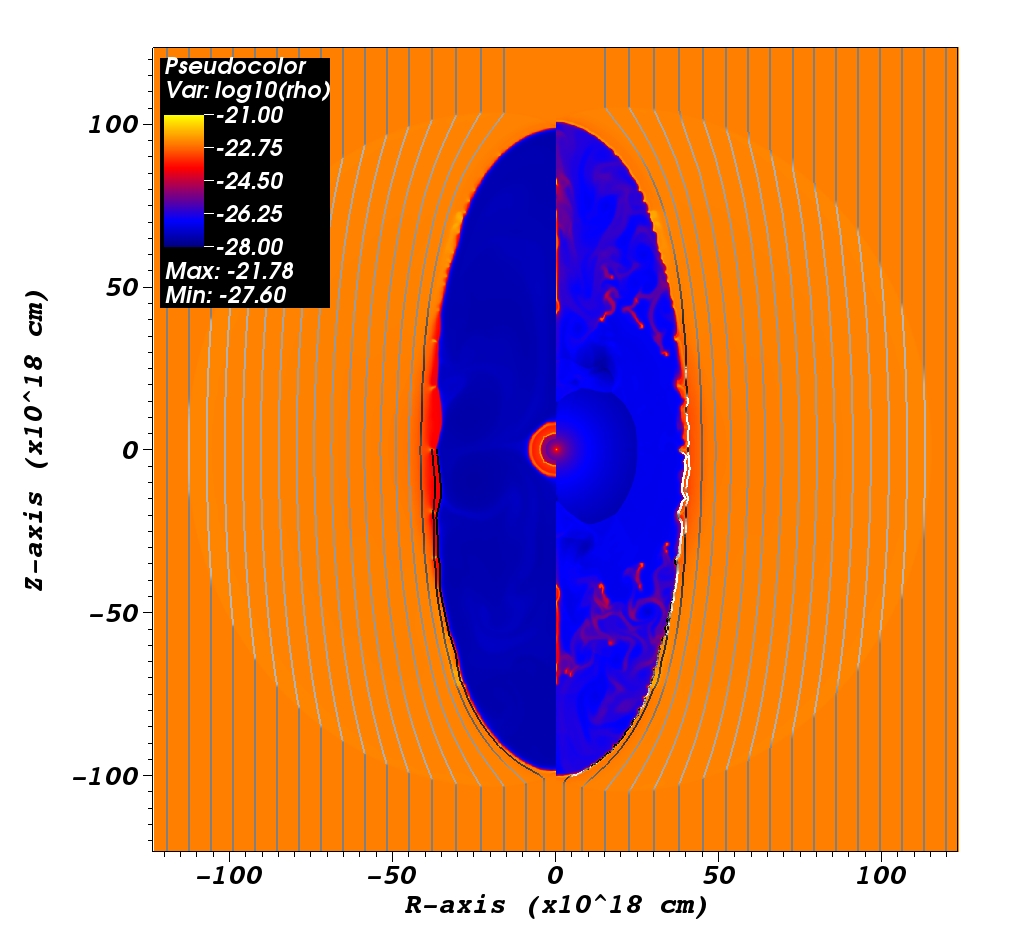}}}
}
\caption{Similar to Figs.~\ref{fig:wsnB0_fig1} and \ref{fig:wsnB5_fig1} but for Simulation~C with an external magnetic field of 10\muG. 
This plot shows both the logarithm of the density in [cgs] and the magnetic field lines. 
The scale of the central and right panels in this figure is 80$\times$80\,pc. 
As for simulation~B, the swept-up shell of shocked ISM is relatively thick, 
 which prevents the formation of thin-shell instabilities; and the contact discontinuity between shocked wind and swept-up ISM is strongly asymmetric. 
Although the RSG shell can form freely against the spherically symmetric wind termination shock (right side of centre panel), 
the expanding WR nebula is strongly affected by the asymmetry of the bubble (left side of right panel). 
}
\label{fig:wsnB10_fig1}
\end{figure*}

\begin{figure*}
\FIG{
 \centering
\mbox{
\subfigure
{\includegraphics[width=0.33\textwidth]{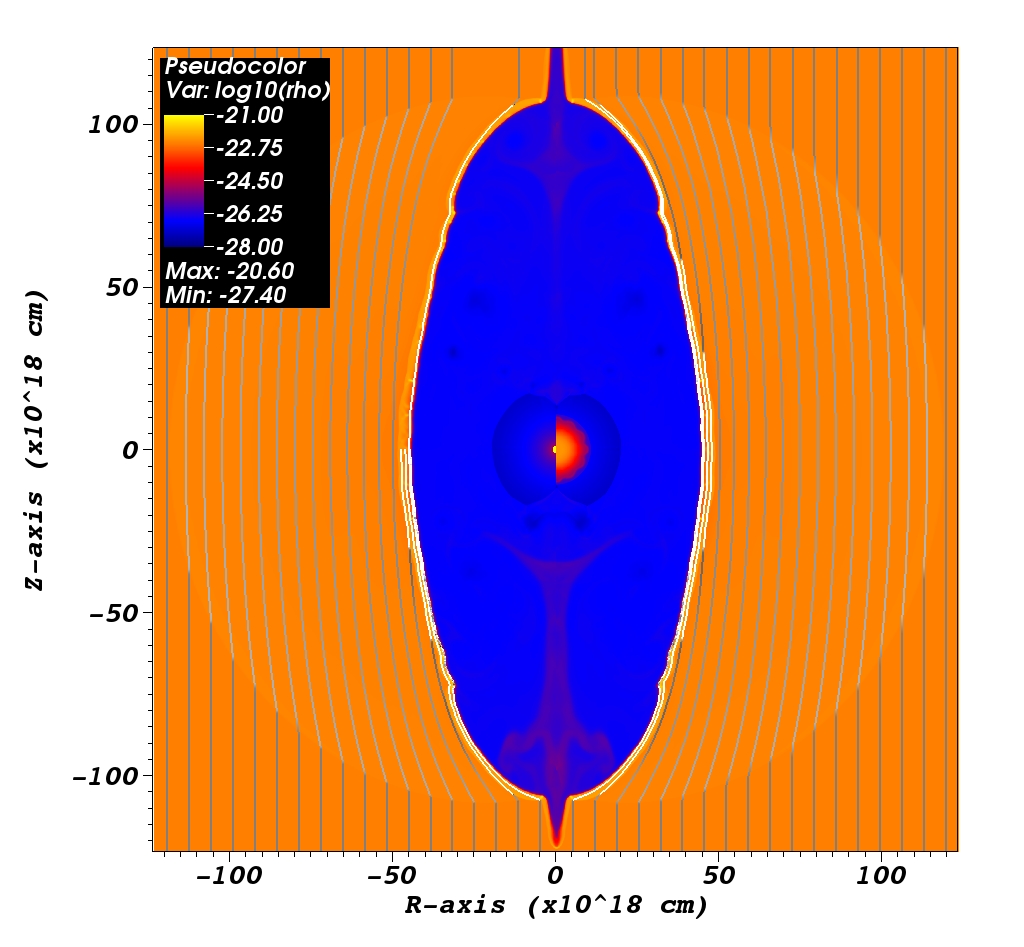}}
\subfigure
{\includegraphics[width=0.33\textwidth]{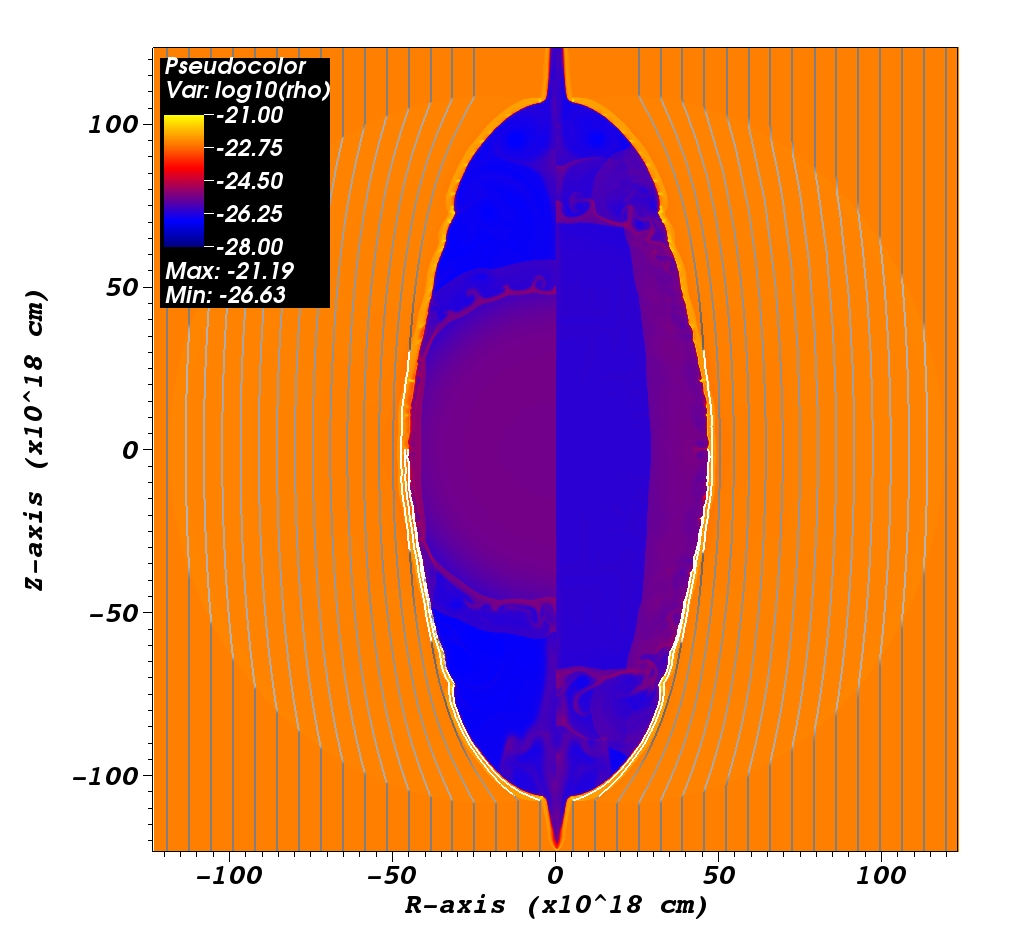}}
\subfigure
{\includegraphics[width=0.33\textwidth]{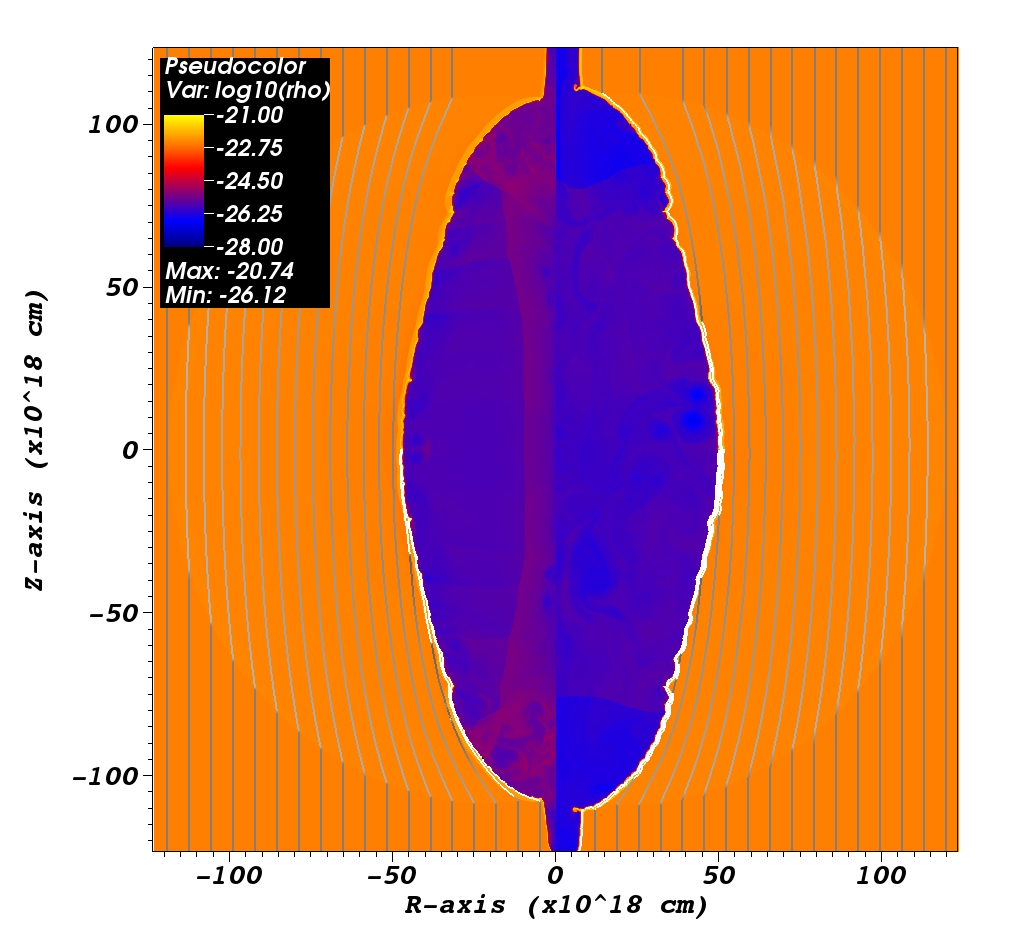}}}
}
\caption{Similar to Figs.~\ref{fig:wsnB0_fig2} and \ref{fig:wsnB5_fig2} but for simulation~C with an external magnetic field of 10\muG. 
Because the supernova expansion is spherical and the outer boundary of the bubble is strongly ellipsoid, 
the supernova will first collide with the outer shell at the ``waist'' of the bubble, noticeable on the left side of the right panel. 
In fact, on the right side of the right panel, the supernova is already moving back toward the centre in the perpendicular direction, 
whereas it is still expanding in the direction parallel to the field.
}
 \label{fig:wsnB10_fig2}
\end{figure*}

\subsection{Intermediate interstellar magnetic field: 10\muG}
During the main sequence evolution, Simulation~C, which uses a magnetic field of 10\muG~\citep[high for the spiral arms, 
but low for the galactic bulge according to][]{RandKulkarni:1989,OhnoShibata:1993,Shabalaetal:2010}, 
shows a similar evolution as Simulation~B, though with a more asymmetrical bubble. 
Initially, the difference between the 5\muG\,and 10\muG\,fields is barely noticeable (extreme left of Fig.~\ref{fig:wsnB10_fig1}). 
However, as time progresses the greater strength of the 10\muG\,field becomes clear (right side of left panel of Fig.~\ref{fig:wsnB10_fig1}),  
leading to a bubble with an asymmetry at the contact discontinuity of approximately 2.5:1 
at the end of the main sequence (left side of centre panel in Fig.~\ref{fig:wsnB10_fig1}). 
Along the longer axis, the radius of the bubble is approximately 31\,pc, which makes it slightly larger than the radius of the bubble in an ISM lacking a magnetic field. 
Along the shorter axis the bubble has a radius of only 12\,pc.
As for Simulation~B, the outer edge of the shell 
(visible only as a slight distortion of the magnetic field lines) remains nearly spherical 
because of the pressure exerted by the compressed magnetic field. 

The strong asymmetry of the bubble influences the evolution of the circumstellar nebula formed during the WR phase (right panel of Fig.~\ref{fig:wsnB10_fig1}). 
In the direction perpendicular to the magnetic field the nebula collides with the swept-up ISM shell much earlier than in the direction parallel to the field. 
As a result, material is driven back toward the central axis where it piles 
up\footnote{This effect is exacerbated by the 2D symmetry of the simulations, which does not allow flow over the poles.}. 

Once the supernova explodes, the  asymmetry of the outer bubble strongly influences its expansion. 
In the direction perpendicular to the field, 
the supernova remnant hits the outer edge of the shocked wind bubble early (left side of centre panel in Fig.~\ref{fig:wsnB10_fig2}). 
Even as it is already moving back toward the centre from this collision, 
it is still expanding in the direction parallel to the field (right side of centre panel in Fig.~\ref{fig:wsnB10_fig2}). 
Eventually, the returning supernova expansion collides with itself, along the major axis of the bubble. 
By this time most of its momentum has been lost and its velocity becomes subsonic (right panel of Fig.~\ref{fig:wsnB10_fig2}).

\begin{figure*}
\FIG{
 \centering
\mbox{
\subfigure
{\includegraphics[width=0.33\textwidth]{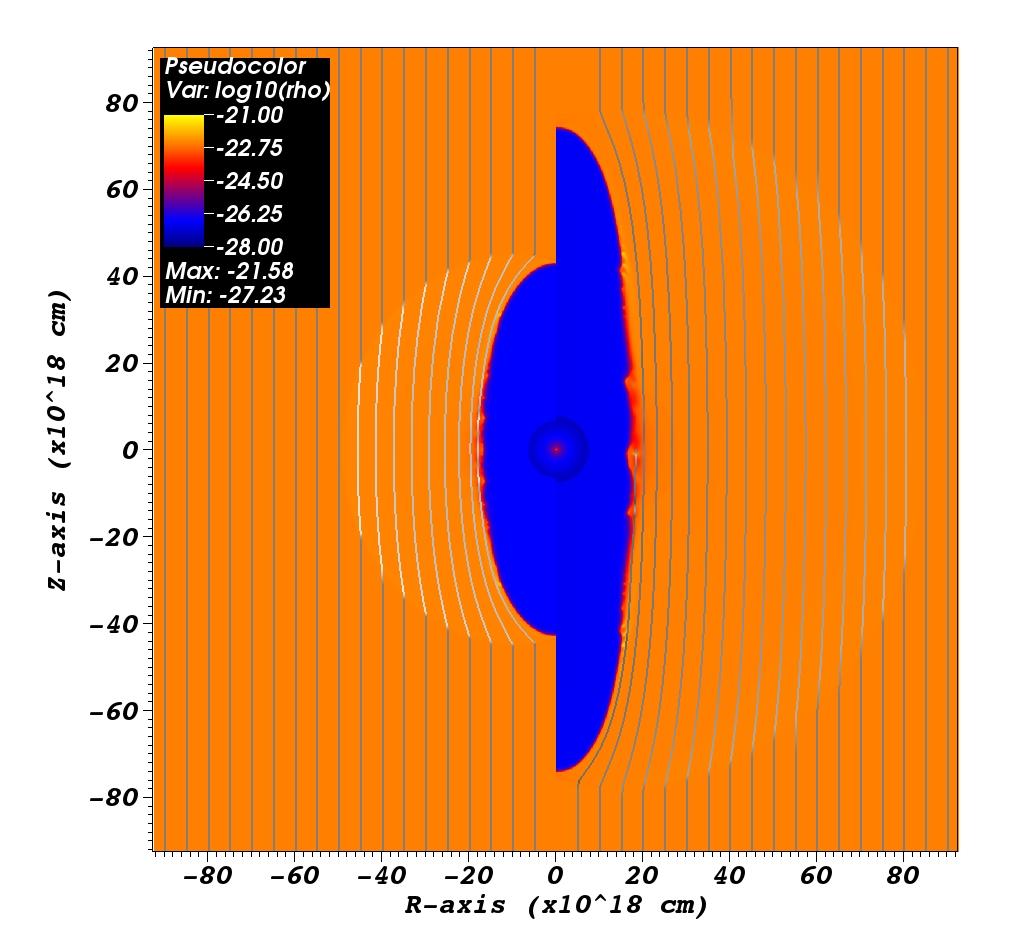}}
\subfigure
{\includegraphics[width=0.33\textwidth]{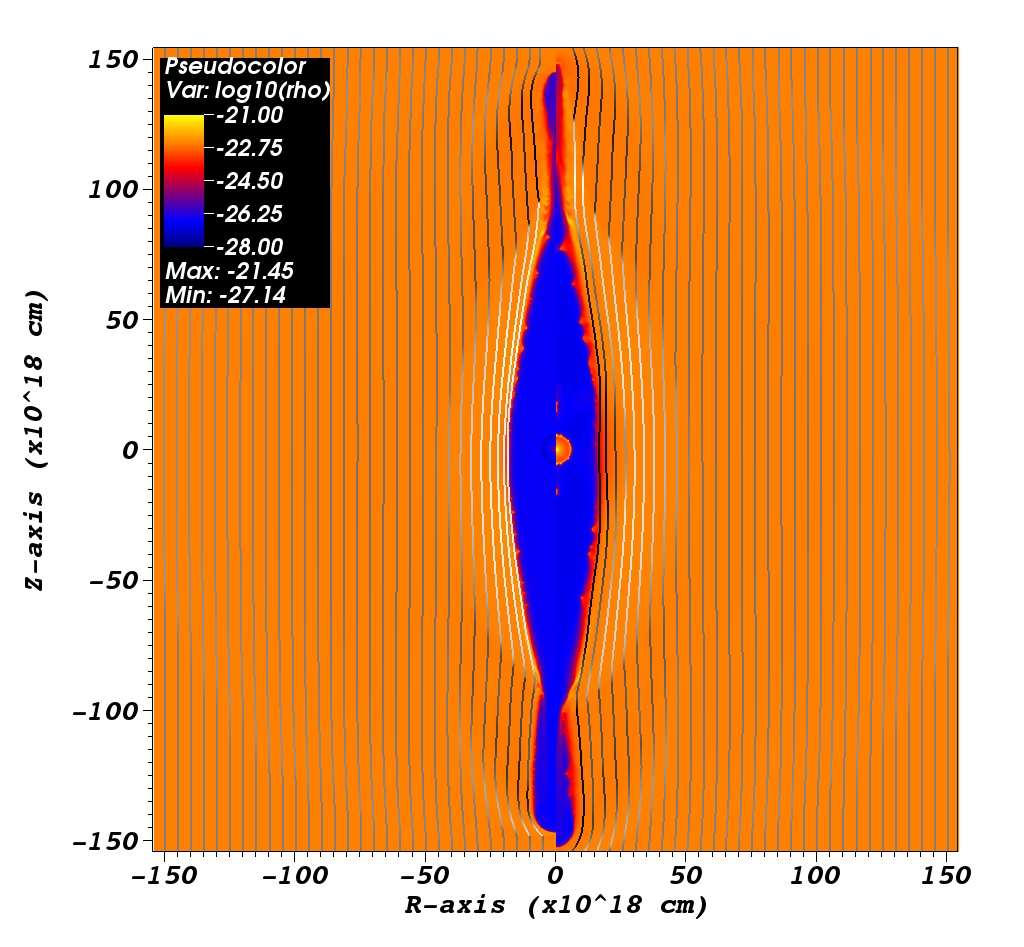}}
\subfigure
{\includegraphics[width=0.33\textwidth]{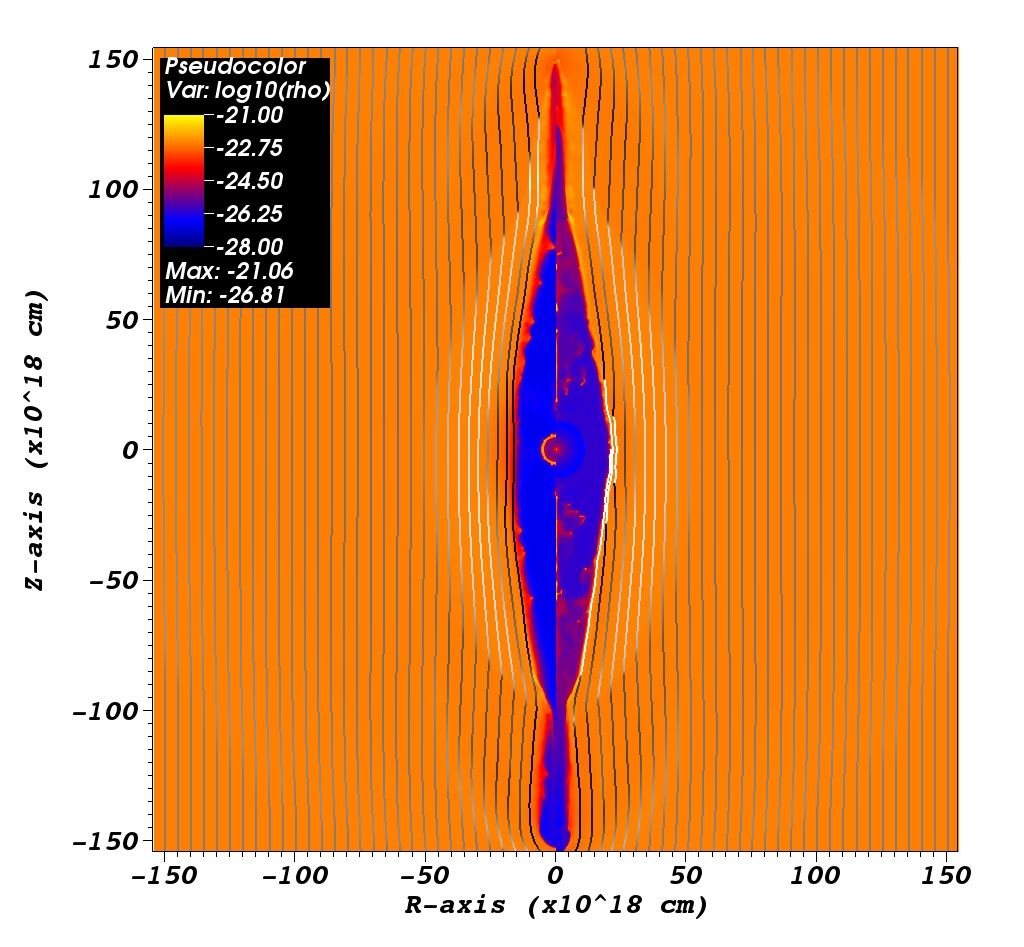}}}
}
\caption{Similar to Figs.~\ref{fig:wsnB0_fig1}, \ref{fig:wsnB5_fig1} and \ref{fig:wsnB10_fig1}, but for simulation~D, which has a 20\muG\, interstellar magnetic field 
The shocked wind bubble is highly asymmetric from the earliest stages of the evolution and stops expanding in the direction perpendicular to the magnetic field 
after it reaches approximately 7\,pc. The elongated shape of the bubble strongly influences the development of the WR nebula (right panel).}
\label{fig:wsnB20_fig1}
\end{figure*}

\begin{figure*}
\FIG{
 \centering
\mbox{
\subfigure
{\includegraphics[width=0.33\textwidth]{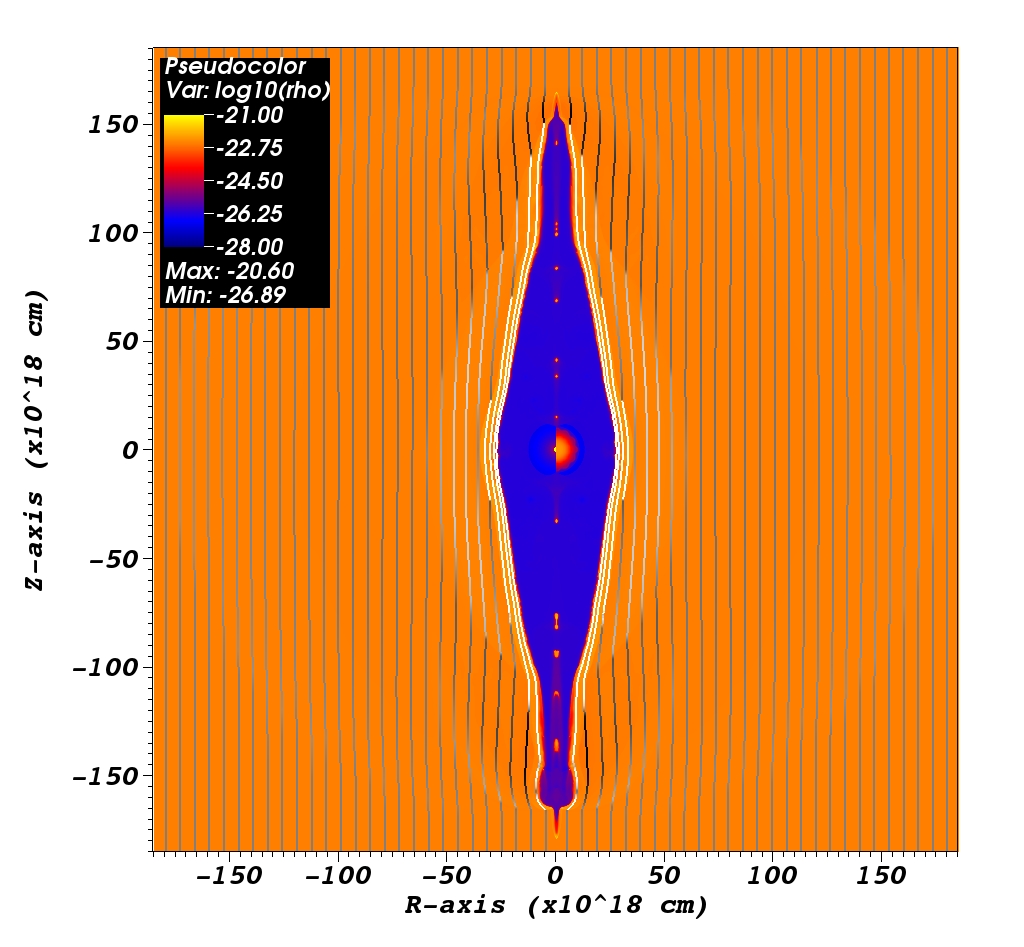}}
\subfigure
{\includegraphics[width=0.33\textwidth]{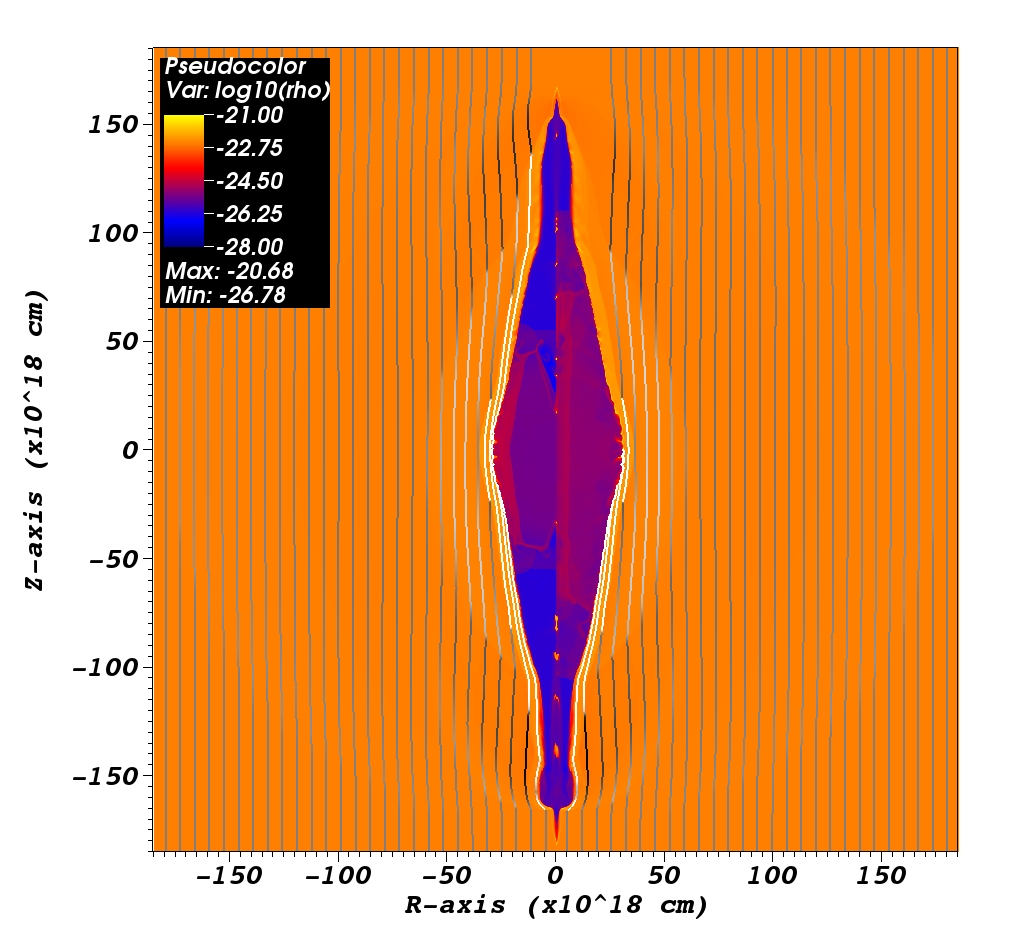}}
\subfigure
{\includegraphics[width=0.33\textwidth]{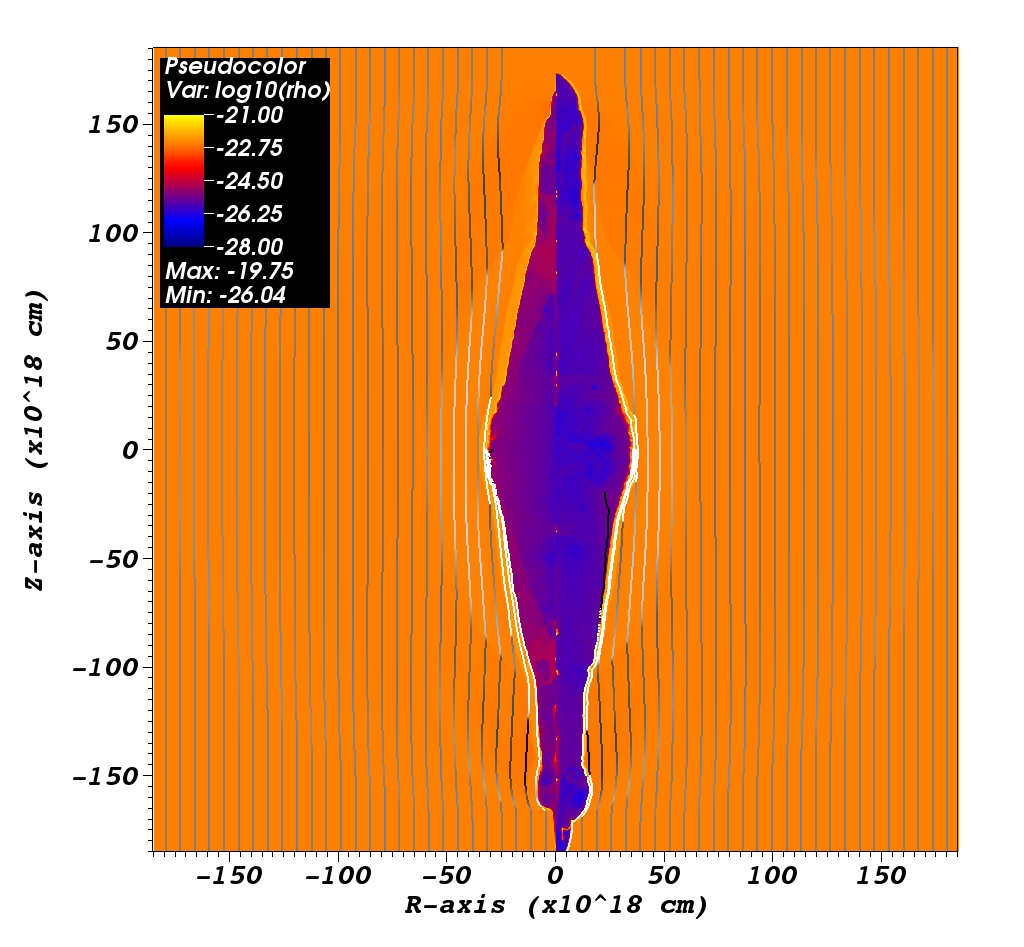}}}
}
\caption{Similar to Figs.~\ref{fig:wsnB0_fig2}, \ref{fig:wsnB5_fig2} and \ref{fig:wsnB10_fig2} but for simulation~D with an external magnetic field of 20\muG. 
The scale of each panel is 120$\times$120\,pc. 
The strongly asymmetrical shape of the bubble causes the supernova to expand along the z-axis after recoiling from the outer edge. 
}
 \label{fig:wsnB20_fig2}
\end{figure*}

\subsection{High interstellar magnetic field: 20\muG}
Because of the external magnetic field strength, which, at 20\muG, is four times as high as in simulation~B, simulation~D has from an early stage a much more asymmetric bubble 
(left panel of Fig.~\ref{fig:wsnB20_fig1}). 
In the direction perpendicular to the magnetic field, the shocked wind bubble is unable to expand beyond about 7\,pc. 
Once it is has reached this point the expansion stops (at least as far as the contact discontinuity between the hot bubble and the swept-up shell is concerned), 
because the thermal pressure in the hot bubble drops below the magnetic pressure of the ISM. 
This also limits the expansion of the wind termination shock, which reaches its maximum distance of 2.7\,pc (right side of left panel of Fig.~\ref{fig:wsnB20_fig1}).
Parallel to the magnetic field, the expansion continues, making the shocked wind bubble increasingly a-spherical. 
As in the simulations~B and C, the interstellar magnetic field is disturbed in a sphere around the shocked wind bubble, 
although not as outspoken as a swept-up shell.  

As the bubble becomes increasingly a-spherical, a pressure gradient starts to form along the major axis, 
with the thermal pressure in the ``tips'' of the bubble becoming lower than the ram pressure of the wind at the termination shock. 
This gives the bubbles a somewhat eye-like shape, reminiscent of the much smaller 
eye-like structures observed around some Asymptotic Giant Branch stars \citep{Coxetal:2012}. 
To the ``north'' of the star, the magnetic field actually squeezes off part of the bubble (centre panel of Fig.~\ref{fig:wsnB20_fig2}), 
though it should be noted that this effect is enhanced by the 2.5-D symmetry of the model.  
As a result of the reduced effective volume of the bubble, 
the thermal pressure increases temporarily, pushing the wind termination shock back toward the star. 
The resulting bubble shape is reminiscent of the ``eye-like'' shapes found for AGB stars by \citet{vanMarleetal:2014b}, albeit on a much larger scale.

When the star becomes a RSG, the ram-pressure of the wind decreases considerably 
(right side of centre panel in Fig.~\ref{fig:wsnB20_fig1}). 
Since the cross-section of the bubble perpendicular to the magnetic field is much shorter than in the direction parallel to the field,  
the effect of change in the wind is first noticeable around the ``waist'' of the bubble. 
The reduced ram-pressure leads to a reduction in thermal pressure. 
This in turn, causes the contact discontinuity to move back toward the star when the thermal pressure of the shocked wind can no longer counterbalance the magnetic pressure of the ISM. 
The reverse happens when  the star reaches the WR phase and the WR shell sweeps up the free-streaming RSG wind as well as the RSG shell at the termination shock. 
Unable to expand in the r-direction, 
the shells collide with the outer edge of the shocked wind bubble and are focussed along the longitudinal axis of the bubble, forming a jet-like structure. 
Afterwards, the strong WR wind causes an increase in thermal pressure in the shocked wind bubble.  
Once again, this is first felt at the ``waist'' of the bubble, which now expands until the magnetic field brings it to a halt again. 
The increased thermal pressure also re-opens the connection to the squeezed-off part (left side of left panel of Fig.~\ref{fig:wsnB20_fig2}). 
At the end of the stellar evolution, the bubble is still highly a-spherical, with  an asymmetry of approximately 7:1. 
Along the axis parallel to the magnetic field, the radius is 50\,pc, but at the ``waist'' the radius is only about 7\,pc.

Once the supernova explodes, the expanding remnant quickly collides with the outer shell because of the narrow ``waist'' of the bubble 
and recoils, flowing back to the central axis. 
It then expands along the length (centre panel of Fig.~\ref{fig:wsnB20_fig2}). 
Eventually, the expansion reaches the end of the elongated bubble and recoils toward the centre, filling up the cavity (right panel of Fig.~\ref{fig:wsnB20_fig2}).

\begin{figure*}
\FIG{
 \centering
\mbox{
\subfigure
{\includegraphics[width=0.33\textwidth]{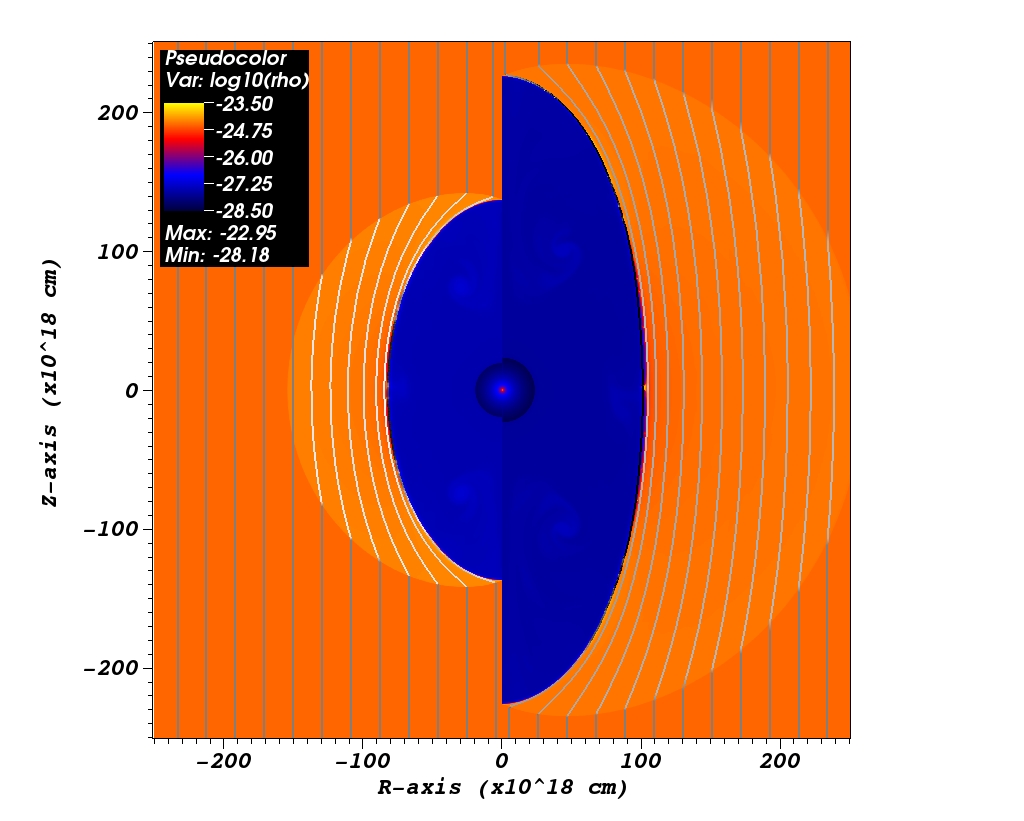}}
\subfigure
{\includegraphics[width=0.33\textwidth]{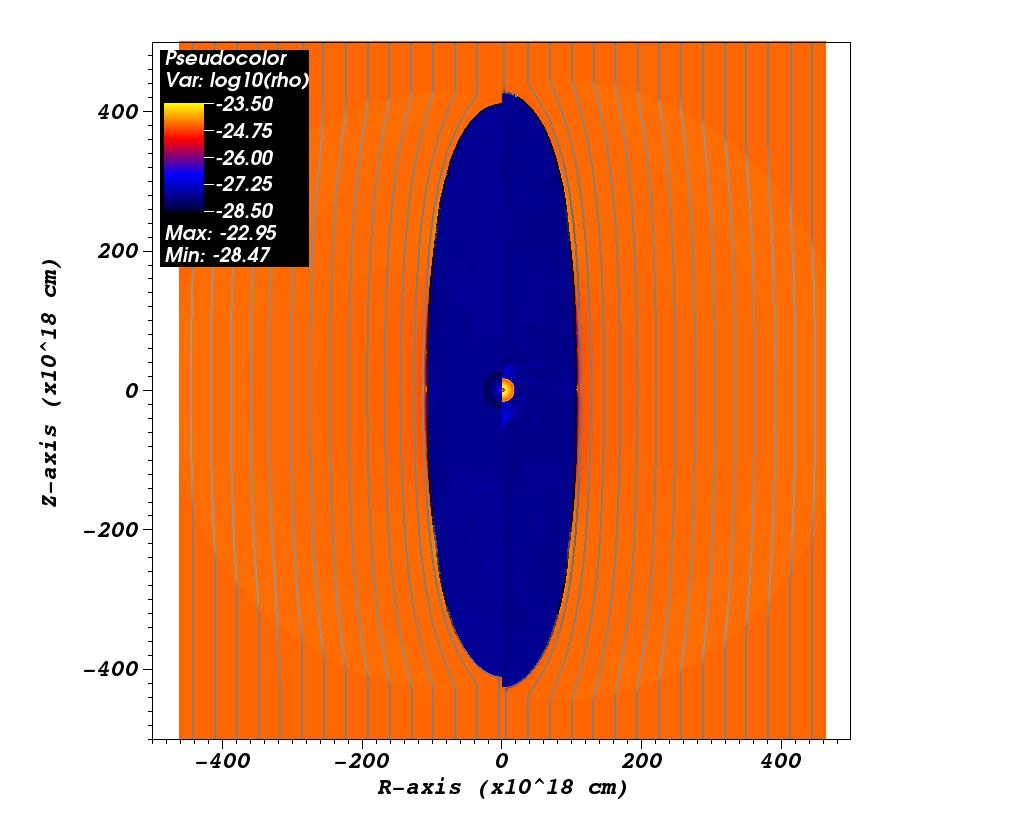}}
\subfigure
{\includegraphics[width=0.33\textwidth]{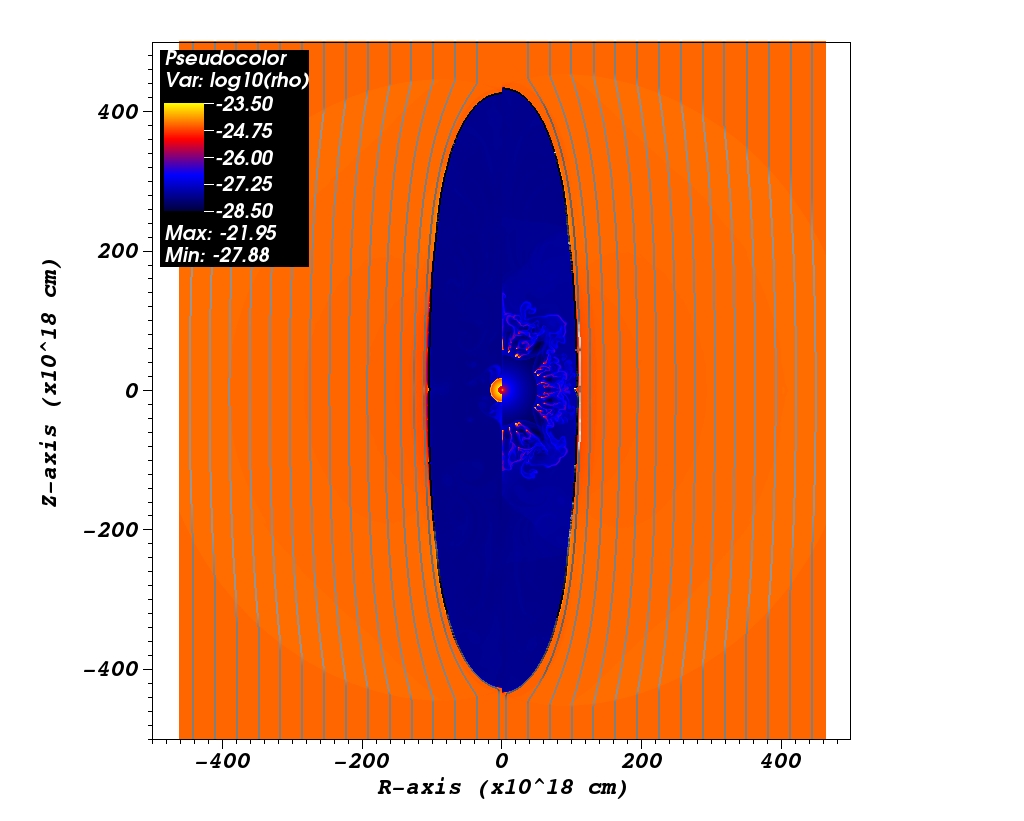}}}
}
\caption{ Similar to Figs.~\ref{fig:wsnB0_fig1}, \ref{fig:wsnB5_fig1}, \ref{fig:wsnB10_fig1} and \ref{fig:wsnB20_fig1}, but for simulation~E, 
which combines a 5\muG\, magnetic field with a warm, low density medium. 
The bubble expands very quickly and becomes much larger than for the previous simulations. 
(Note the different scale: the left panel is 160 by 160 pc, the centre and right panel are 290 by 290 pc.) 
This allows the red supergiant and Wolf-Rayet shells to form 
unimpaired by the aspherical shape of the bubble.}
\label{fig:wsnB05lowdens_fig1}
\end{figure*}

\begin{figure*}
\FIG{
 \centering
\mbox{
\subfigure
{\includegraphics[width=0.33\textwidth]{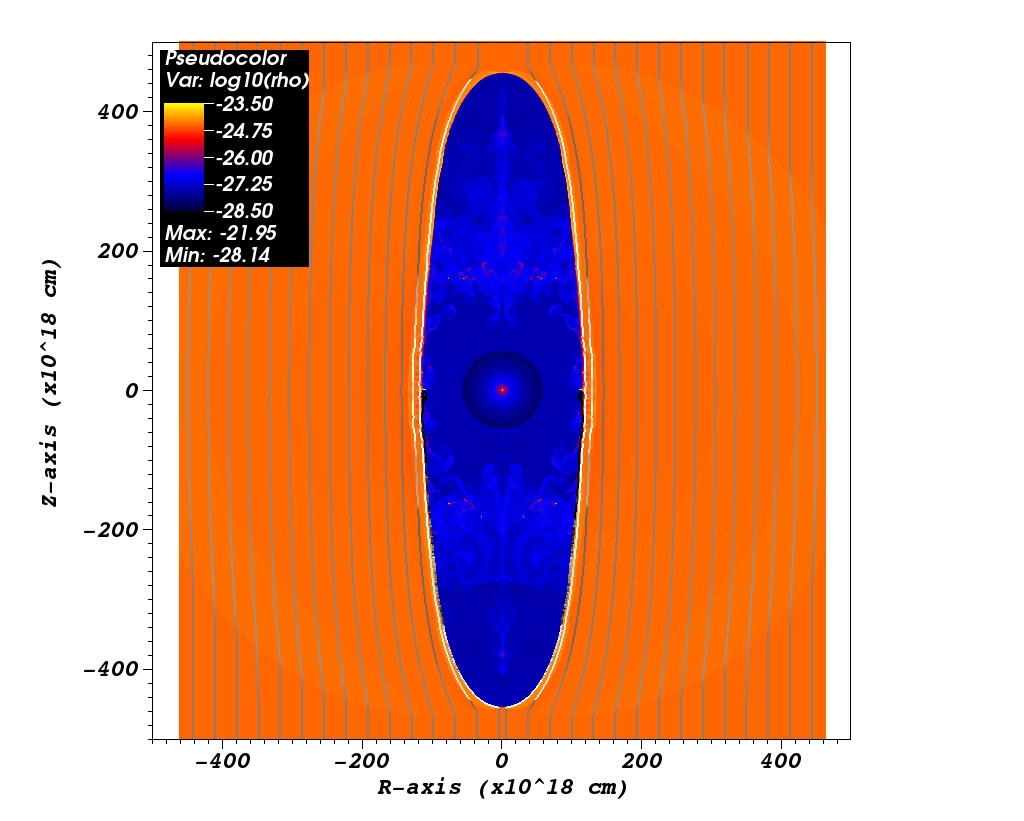}}
\subfigure
{\includegraphics[width=0.33\textwidth]{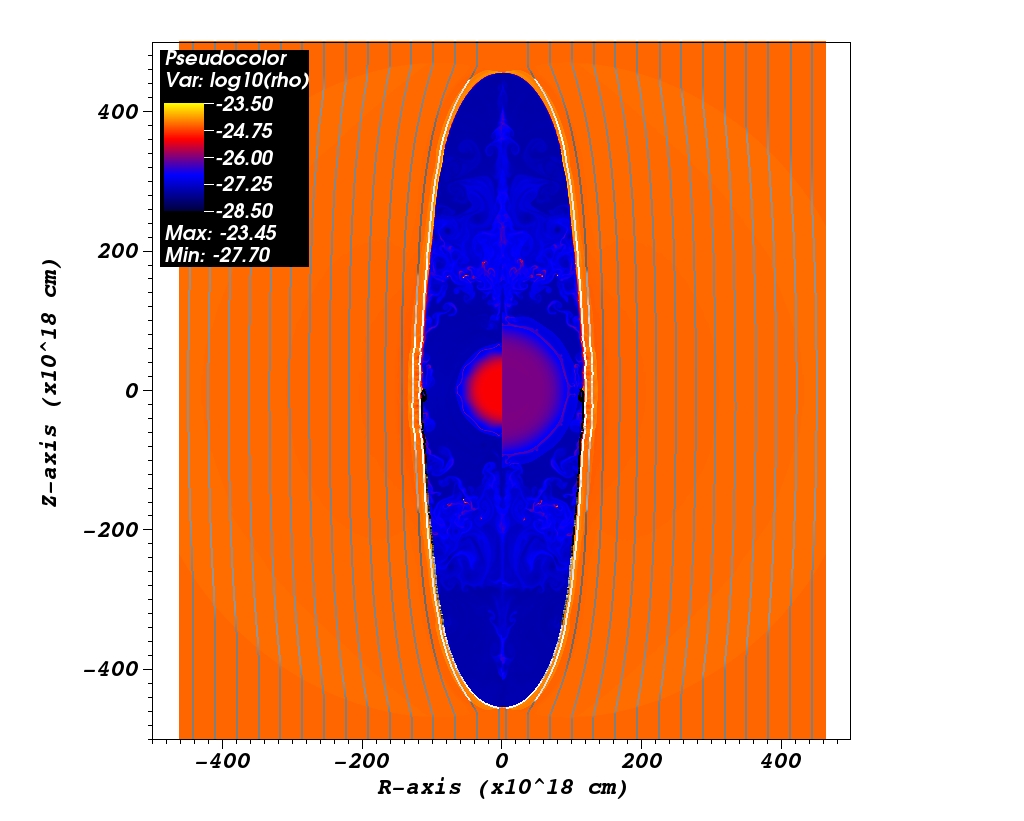}}
\subfigure
{\includegraphics[width=0.33\textwidth]{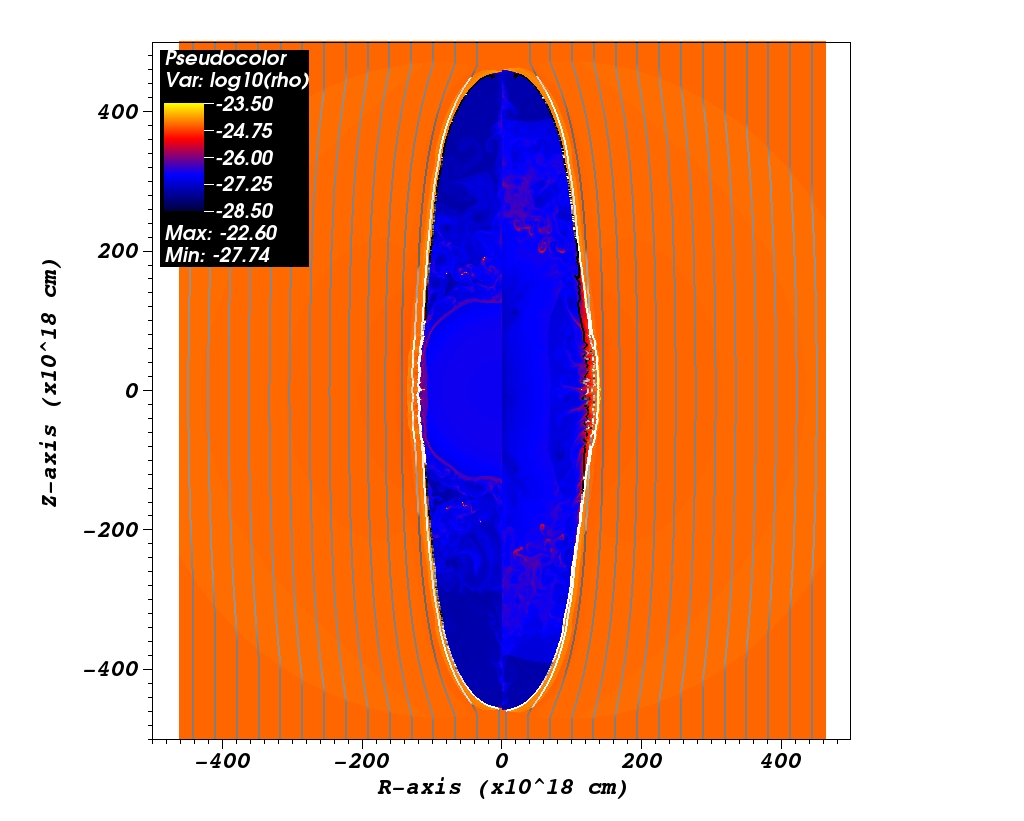}}}
}
\caption{ Similar to Figs.~\ref{fig:wsnB0_fig2}, \ref{fig:wsnB5_fig2},  \ref{fig:wsnB10_fig2} and \ref{fig:wsnB20_fig1} but for simulation~E, 
which combines a 5\muG\, magnetic field with a warm, low density medium. All panels are 290 by 290 pc.  
The impact of the supernova remnant with the contact discontinuity causes the local gas to be compressed, which leads to the formation of instabilities (right panel).
}
 \label{fig:wsnB05lowdens_fig2}
\end{figure*}

\section{Bubbles in a warm ISM}
\label{sec-warm}
\subsection{The influence of a warm, low density ISM}
Simulation~E (Figs.~\ref{fig:wsnB05lowdens_fig1}-\ref{fig:wsnB05lowdens_fig2}) combines a warm IMS (10\,000\,K) with low density (10$^{-24.5}\gcm$) and a 5\muG, magnetic field. 
Because both the ISM thermal pressure and the ISM inertia are low, owing to the low density, 
the bubble can expand very quickly (left panel of Fig.~\ref{fig:wsnB05lowdens_fig1}), remaining supersonic for the early part of the main sequence phase, 
despite the higher sound speed in the ISM.  
This leaves the magnetic field as the dominant force in the ISM, though it is insufficient to stop the expansion. 
The end result is a very large, ellipsoid bubble.
By the end of the main sequence (left side, centre panel of Fig.~\ref{fig:wsnB05lowdens_fig1}), the bubble has reached a size of approximately 
270 by 38 parsec.    
Because of the extremely large size of the bubble, the development of the internal shells (RSG shell and Wolf-Rayet ring nebula) proceeds unimpaired, 
with the Wolf-Rayet ring nebula retaining its spherical shape and breaking up before it 
collides with the outer edge of the main sequence bubble (right panel of Fig.~\ref{fig:wsnB05lowdens_fig1}). 

The supernova expansion also proceeds largely uninhibited (left and centre panels of Fig.~\ref{fig:wsnB05lowdens_fig2}). 
However, unlike the previous simulations (A-D), the expansion of the supernova remnant actually leaves a visible 
imprint on the interface with the ISM, creating a thin, localized shell and some deformation of the bubble (right panel of Fig.~\ref{fig:wsnB05lowdens_fig2}). 
This is the result of the low ISM density, which allows the contact discontinuity to accelerate outward under the force excerted by the collision with the supernova remnant, 
compressing the interestellar gas into a thin shell, despite the magnetic field pressure. 
This also leads to the formation of Rayleigh-Taylor instabilities at the contact discontinuity (right side of right panel of Fig.~\ref{fig:wsnB05lowdens_fig2}).

\begin{figure*}
\FIG{
 \centering
\mbox{
\subfigure
{\includegraphics[width=0.5\textwidth]{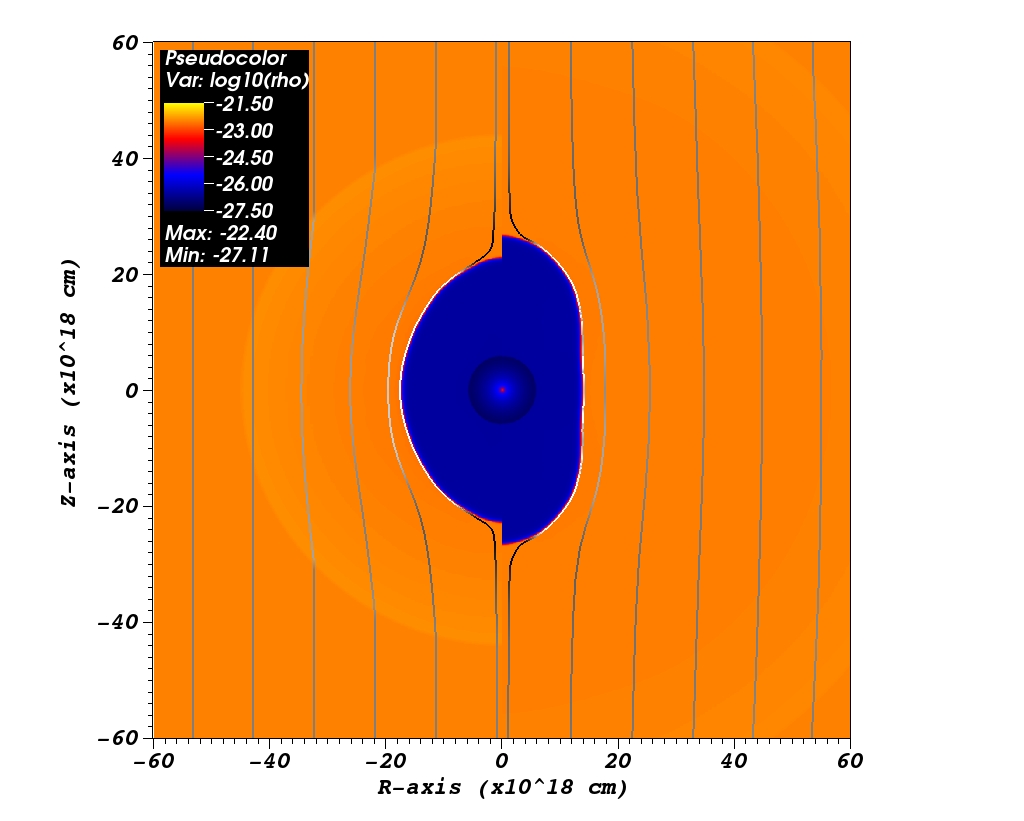}}
\subfigure
{\includegraphics[width=0.5\textwidth]{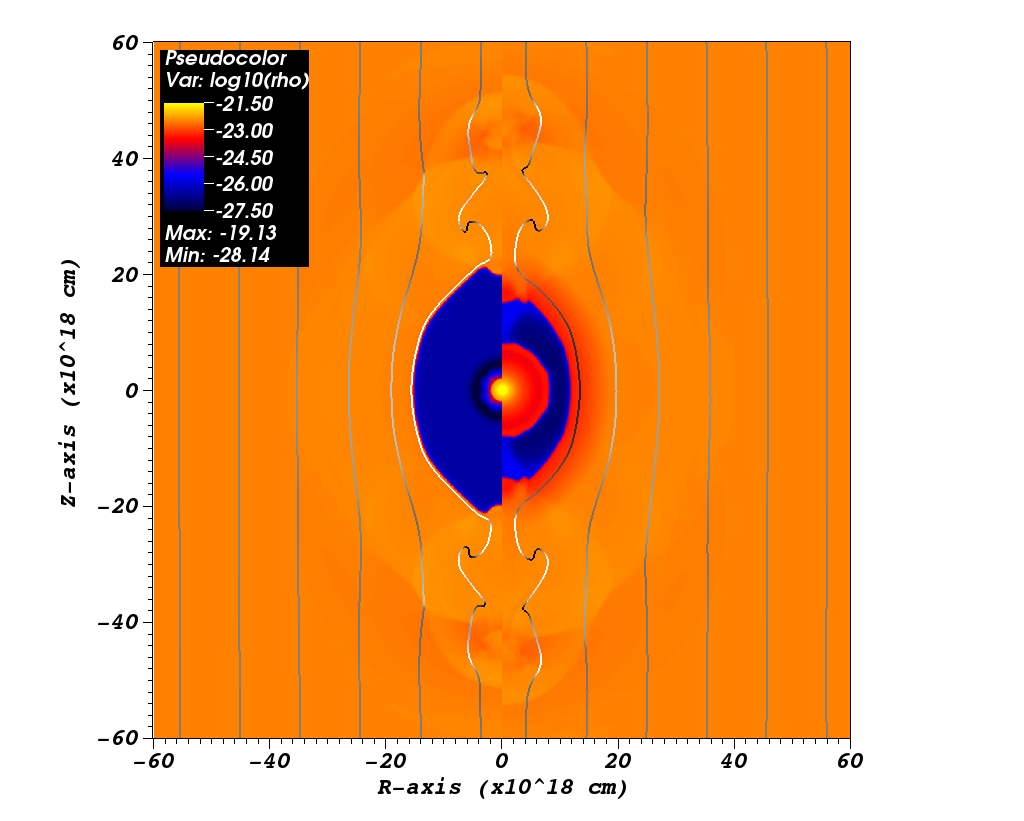}}}
}
\caption{  Development of a circumstellar bubble during the main sequence in a high density, warm ISM (simulation~F) with a 5\muG\, interstellar magnetic field. 
From left to right this figure shows density and magnetic field lines at 1,2, 4.3, and 4.5\,Myr. 
The bubble is much smaller compared to the other 5\muG\, simulations (B $\&$ E). 
The combined thermal and magnetic pressure in the ISM is strong enough to reduce the size of the bubble considerably. 
By the end of the main sequence (left side of right panel), the bubble has obtained a somewhat ``eye-like'' shape. 
At the onset of the red supergiant phase (right side of right panel), 
the hot, shocked wind bubble collapses under the combined
forces of the interstellar magnetic field and the thermal pressure in the ISM. 
}
\label{fig:wsnB5warm_fig3}
\end{figure*}

\begin{figure*}
\FIG{
 \centering
\mbox{
\subfigure
{\includegraphics[width=0.5\textwidth]{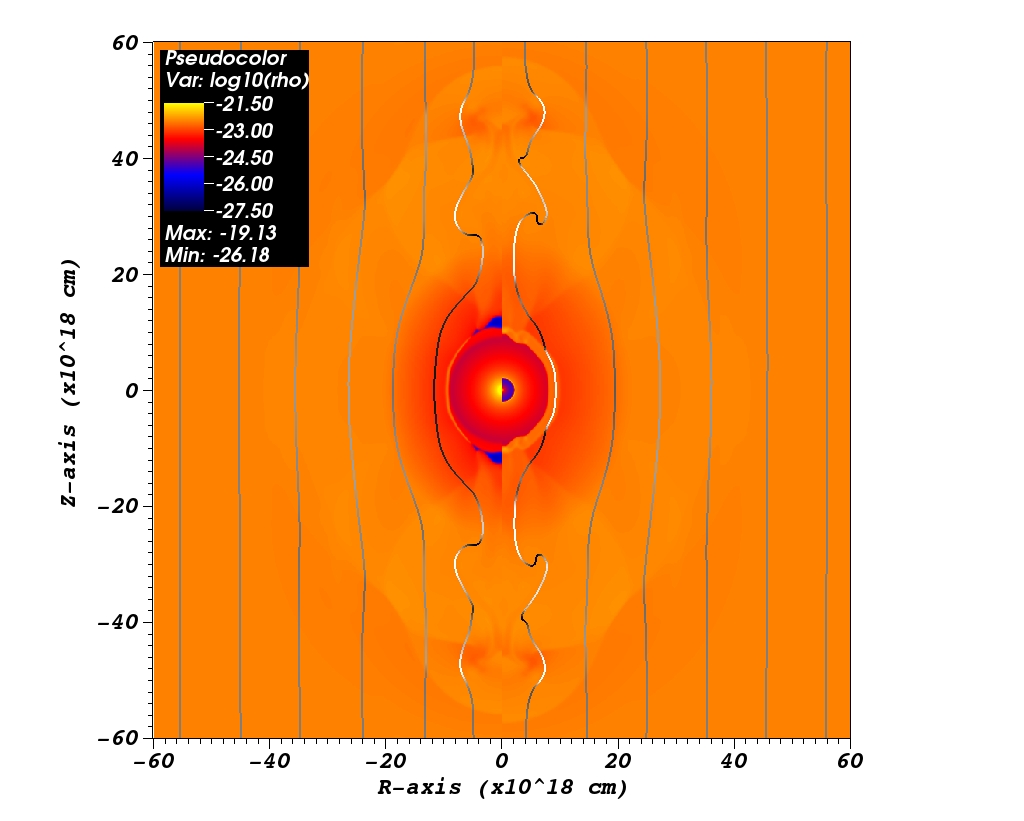}}
\subfigure
{\includegraphics[width=0.5\textwidth]{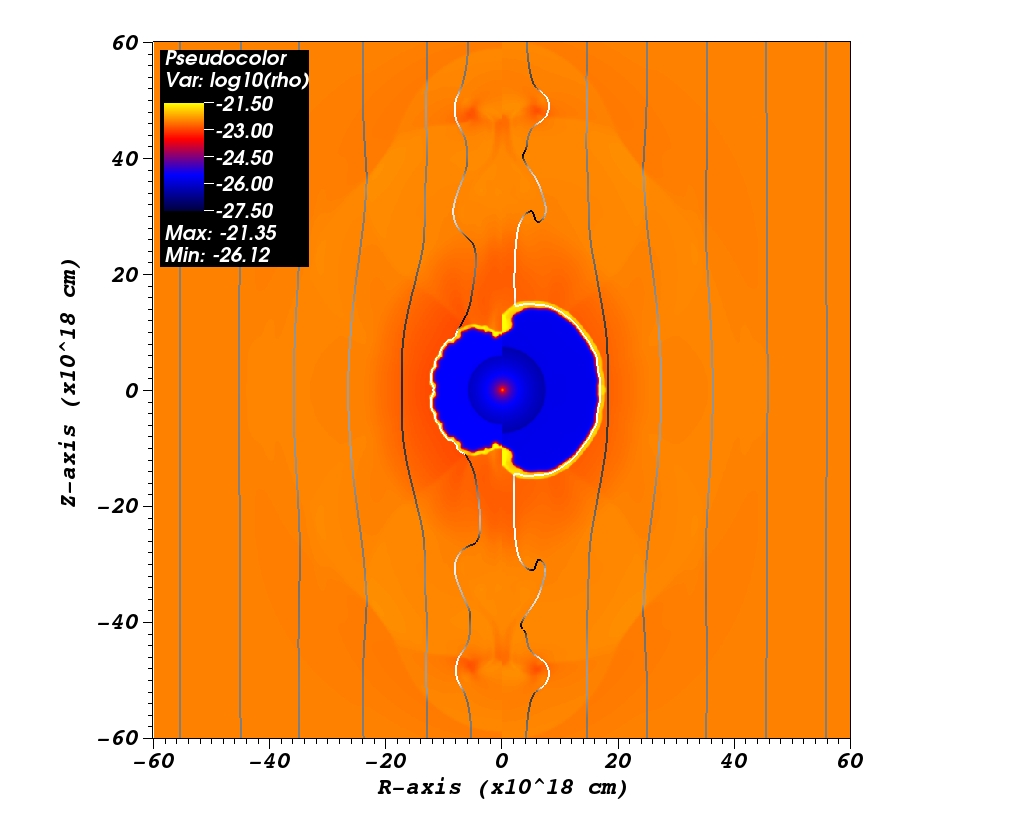}}}
}
\caption{ The shape of the bubble for simulation~F after 4.45, 4.5, 4.55 and 4.6~Myr. 
During the red supergiant phase (left panel), 
the hot, shocked wind bubble is destroyed completely by the combined
forces of the interstellar magnetic field and the thermal pressure in the ISM. 
The result is a purely radiative collision between the red supergiant wind and the outer shell (left side of left panel.) 
Once the start reaches the Wolf-Rayet stage (rightside, left panel), the powerful wind quickly sweeps up the red supergiant wind, 
forming a new hot bubble that starts to expand into the ISM (right panel). 
}
\label{fig:wsnB5warm_fig4}
\end{figure*}

\begin{figure*}
\FIG{
 \centering
\mbox{
\subfigure
{\includegraphics[width=0.33\textwidth]{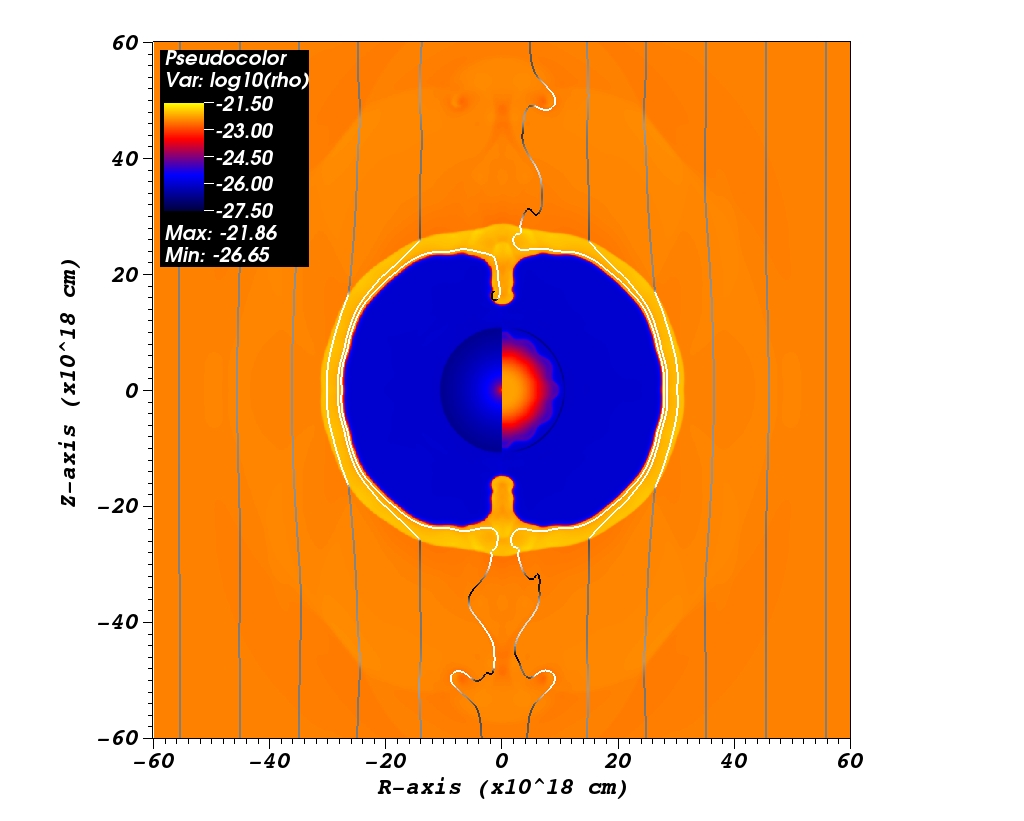}}
\subfigure
{\includegraphics[width=0.33\textwidth]{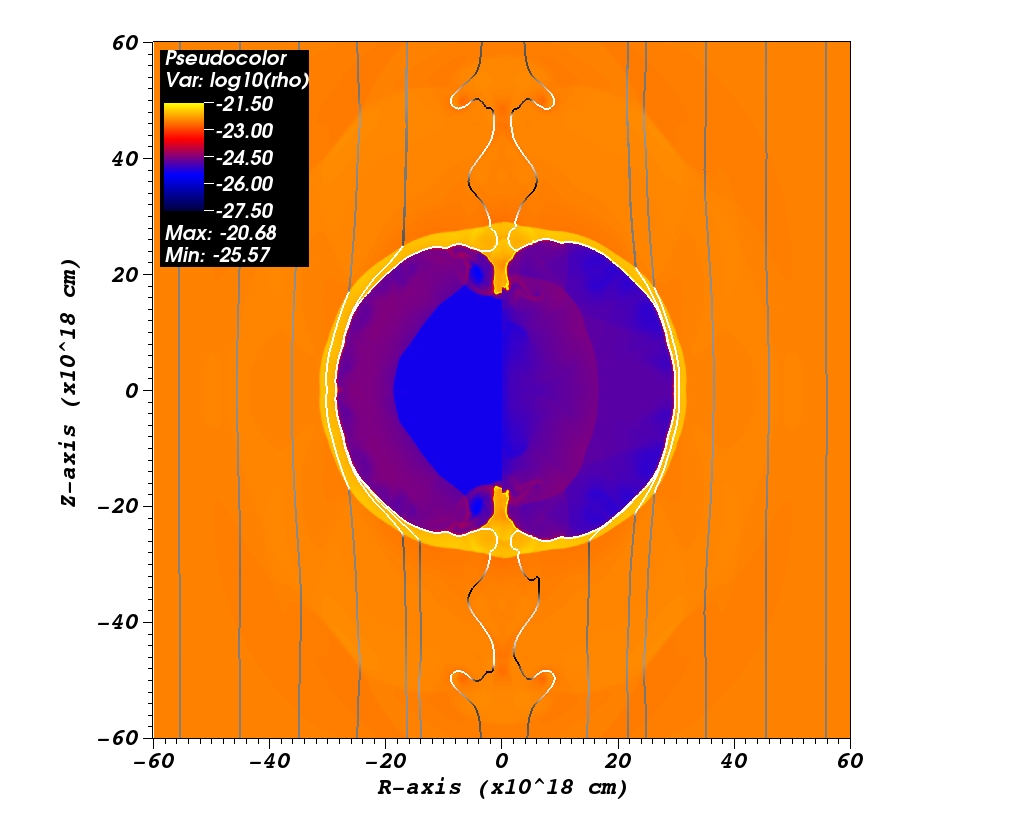}}
\subfigure
{\includegraphics[width=0.33\textwidth]{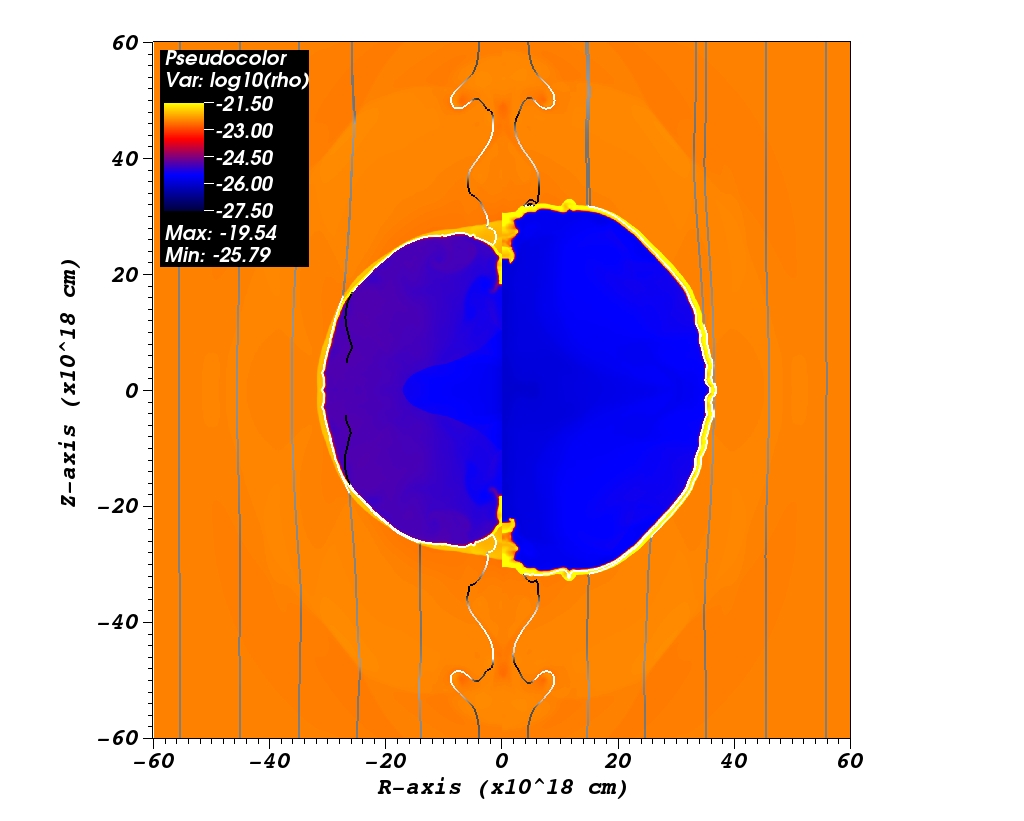}}}
}
\caption{ Similar to Figs.~\ref{fig:wsnB0_fig2}, \ref{fig:wsnB5_fig2}, \ref{fig:wsnB10_fig2}, \ref{fig:wsnB20_fig2}, and \ref{fig:wsnB05lowdens_fig2}, 
but for simulation~F. This figure shows the expansion of a supernova inside a wind bubble that is constrained by both high thermal pressure and magnetic field pressure. 
The supernova is effectively trapped inside a relatively small bubble and, as in all previous models, fails to break out of the shell.
}
\label{fig:wsnB5warm_fig5}
\end{figure*}

\subsection{The influence of a warm, dense ISM}
 For simulation~F, we repeat simulation~B but with an ISM temperature of 10\,000\,K. 
This type of combination (both warm and high density) is expected to reside in
H$_{\rm II}$ regions in the close vicinity of massive stellar clusters. 
It serves to demonstrate 
how the interstellar magnetic field and ISM thermal pressure combined can influence the shape and size of the circumstellar bubble, leading to a radically different bubble evolution.

Under these conditions the thermal pressure in the ISM ($2.6\times10^{-11}$\,dyne/cm$^2$) is more than an order of magnitude higher than the magnetic field pressure. 
The effect of the two pressures combined leads to a highly a-typcial bubble development as shown 
in Figs.~\ref{fig:wsnB5warm_fig3}-\ref{fig:wsnB5warm_fig4}. 
Due to the high thermal pressure in the ISM the expansion is slow, being both sub-sonic and sub-Alfv{\'e}nic, which reduces the internal volume of the bubble. 
This in turn increases the internal density, making radiative cooling more efficient, which reduces the internal pressure 
of the bubble, leading to a similar process as in simulation~D: the outer tips of the expanding bubble are squeezed off by a combination 
of thermal and magnetic pressure from the ISM, resulting in a somewhat ``eye-shaped'' bubble ( left side of right panel of Fig.~\ref{fig:wsnB5warm_fig3}), similar to the ones found around 
some AGB stars \citep{Coxetal:2012,vanMarleetal:2014b}, though it should be noted that this effect may be an artifact of the 2D nature of our simulations (see Section~\ref{sec-limitations}). 
Afterwards the bubble starts to expand once more because the energy from the wind is now contained in a relatively small volume. 
However, before it can regain its former size, the star reaches the end of the main sequence and becomes a red supergiant 
 (right side of right panel of Fig.~\ref{fig:wsnB5warm_fig3}). 
The resulting decrease in wind ram pressure stops the expansion and 
causes the bubble to collapse completely under the outside forces ( right side or right panel of Fig.~\ref{fig:wsnB5warm_fig3} and left panel of Fig.~\ref{fig:wsnB5warm_fig4}).
The result is a fundamental change in the interaction between the wind and the ISM. 
Where before the wind termination shock was adiabatic, it now becomes completely radiative with the free-streaming red supergiant wind impacting directly 
onto the ISM. 
By the time the star evolves from red supergiant to Wolf-Rayet star the hot, shocked wind bubble has completely disappeared ( left panel of Fig.~\ref{fig:wsnB5warm_fig4}). 
The Wolf-rayet wind forms a ring nebula around the star (right side of left panel of Fig.~\ref{fig:wsnB5warm_fig4}), which, driven by the hot, shocked Wolf-Rayet wind, 
quickly expands into the red supergiant wind, 
forming a new, hot shocked wind bubble, which continues to expand outward, effectively recreating the original main sequence bubble. 
The expansion is, once again, somewhat a-spherical, owing to the magnetic field pressure as well as the fact that the new expanding shell encounters piled-up mass along 
the z-axis. 
By this stage the bubble closely resembles the results that \citet{vanMarleetal:2014b} found for a planetary nebula expansion inside 
an AGB wind that had been constrained by an interstellar magnetic field. 

By the time the star explodes as a supernova (Fig.~\ref{fig:wsnB5warm_fig5}) the bubble is still small compared to the other bubbles. 
As a result, the supernova remnant quickly 
reaches the outer shell and bounces back. Because the bubble is not extremely aspherical, the interaction between the supernova remnant and the outer shell 
occurs almost simulteaneously along the entire shell and avoids the strongly asymmetrical shape that occured in simulation~D.

\begin{figure*}
\FIG{
 \centering
\mbox{
\subfigure
{\includegraphics[width=0.5\textwidth]{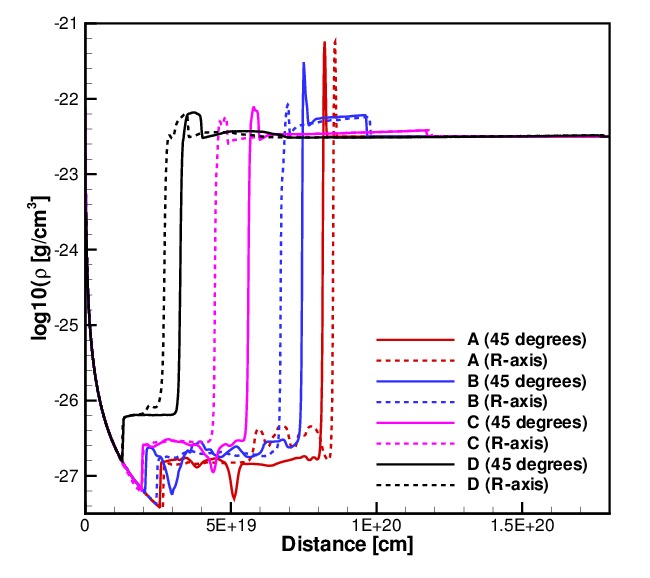}}
\subfigure
{\includegraphics[width=0.5\textwidth]{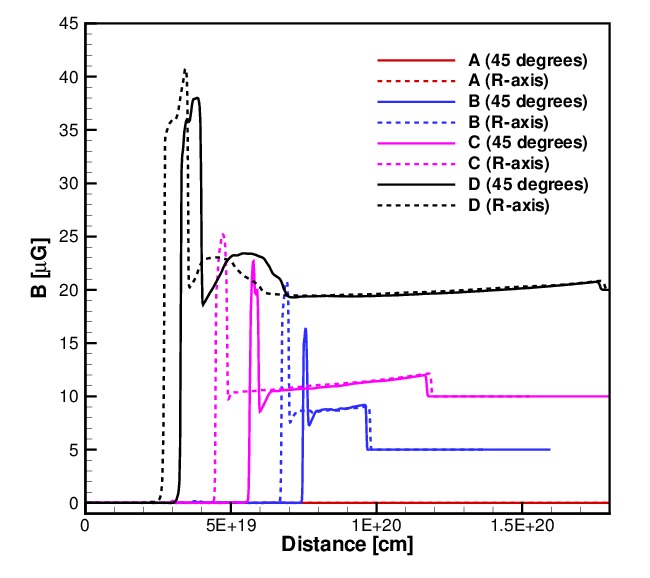}}}
}
\caption{These figures show the gas density (left) and magnetic field strength along cuts through the 2D simulation results at the end of stellar evolution 
(far-right panels of Figs.~\ref{fig:wsnB0_fig2}, \ref{fig:wsnB5_fig2}, \ref{fig:wsnB10_fig2} and \ref{fig:wsnB20_fig2}). 
Each cut starts at the central star and moves either along the R-axis (dashed lines) or at a $45^{\rm o}$ angle (continuous lines) between the R- and Z-axes. 
The compression of the outer shell decreases with increasing field strength, Alfv{\'e}n-waves weakening the shock. 
}
 \label{fig:1Dcuts}
\end{figure*}

\section{Influence of interstellar magnetic fields on the properties of circumstellar bubbles}
\label{sec-discussion}
\subsection{General shape and size of the circumstellar bubble}
As predicted analytically \citep{Heiligman:1980} and shown previously in numerical models by \citet{Tomisaka:1990,Tomisaka:1992} and \citet{Ferriere:1991}, the magnetic field 
limits the expansion of the bubble in the direction perpendicular to the field, partially because of the magnetic tension force, which counteracts the bending of 
the field lines by the expanding bubble, and partially by the magnetic pressure, which counteracts the field lines being compressed. 

A direct comparison between the four bubbles can be made through cross-sections of the bubbles as shown in fig.~\ref{fig:1Dcuts}, 
which show the density (left panel) and magnetic field strength (right panel) along 1-D cuts perpendicular to the field (continuous lines) 
and at a $45^{\rm o}$ angle (dashed lines) for all four simulations at the end of stellar evolution. 
In all cases, the shell is characterized by a thin, highly compressed feature at the contact discontinuity, where the shocked wind directly sweeps up the surrounding medium and an extended, 
only slightly compressed outer shell, where the magnetic field, compressed by the expanding bubble pushes the gas ahead. 
For increasing magnetic field strength, the amount of compression perpendicular to the field in the outer shell decreases, 
until, at a field strength of 20\muG,
one can barely speak of an outer shell. 
This is the consequence of the increase in the Alfv{\'e}n speed, which is defined as,
\begin{equation}
 v_{\rm A}~=~\frac{B}{\sqrt{4\pi\rho}}.
\end{equation}
Even for a 5\muG\, field in an ISM with a density of $10^{-22.5}\gcm$, this gives us an Alfv{\'e}n speed of $2.5\kms$, 
which is significantly higher than the adiabatic sound speed in a cold ISM ($\sim1\kms$ at 100\,K). 
Hence, the strength of the shock is determined by the Alfv{\'e}n speed, rather than the sound speed.  
A higher Alfv{\'e}n speed reduces the strength of the shock and therefore the compression rate. 
At 20\muG, the Alfv{\'e}n speed is approximately $10\kms$.  
At the same time the total expansion of the bubble, perpendicular to the field, at 20\muG\, is only 7\,pc in 4.77 million years: $1.4\kms$. 
Therefore, the expansion perpendicular to the field has become a sub-Alfv{\'e}nic wave, rather than a shock. 
The same effect, though to a lesser extent, is noticeable at a $45^{\rm o}$ angle to the field. 

The behaviour of the bubbles in simulations~B-D generally follows the predictions based on Fig.~\ref{fig:rb_plot}. 
For those circumstances where the ram pressure of the stellar wind is always higher than the magnetic pressure of the interstellar magnetic field, 
as is the case for the 5\muG\, model, 
the bubble becomes ovoid, rather than spherical owing to the magnetic tension force that seeks to reduce the curvature of the field lines, 
but the expansion continues in all directions. 
If the magnetic field pressure exceeds the ram pressure, expansion perpendicular to the field lines ceases altogether, 
resulting in a tube-like bubble 
such as we find for the 20\muG\, model. 
The 10\muG\, model falls in between these two, but generally follows the 5\muG\, model, 
because the ram pressure of the wind only briefly drops below the magnetic field pressure. 

As soon as $P_{\rm ram}< P_{\rm B}$ expansion ceases in the direction perpendicular to the magnetic field, at least for the inner boundary of the swept-up ISM shell. 
This did not occur in the  models by \citet{Tomisaka:1990,Tomisaka:1992} and \citet{Ferriere:1991}, who did not investigate magnetic fields of sufficient strength,  
and in our simulations happens only for the model with $B_{\rm ISM}=20$\muG. 
The outer limit of the shell continues to expand, driven by the magnetic pressure, which reduces the local density inside the shell. 
Interestingly, owing to the extremely wide ``shell'', the field lines are affected to a distance of about 50\,pc, 
despite the fact that the tube-like shocked wind bubble is only about 7\,pc across. 
As calculated in Sect.~\ref{sec-ISM}, for our main sequence wind parameters the radius for the termination shock should be 2.6\,pc 
if the interstellar magnetic field strength is 20\muG. 
This is well in agreement with our numerical results, which put the wind termination shock at a maximum of 2.7\,pc (right side of the left panel of Fig.~\ref{fig:wsnB20_fig1}). 
In theory, the Wolf-Rayet wind is strong enough for the expansion perpendicular to the field lines to resume, even for the 20\muG\, magnetic field. 
We see some sign of this, with the ``waist'' of the bubble expanding. 
However, since the Wolf-Rayet phase is relatively short, it lacks the time to change the general tube-like appearance of the bubble. 
 
As was shown for super bubbles by \citet{Tomisaka:1990,Tomisaka:1992} and \citet{Ferriere:1991}, 
the total volume of the individual bubbles is greatly reduced by the presence of the magnetic ISM field. 
The absolute size of the major axis of the bubble does not change much with the magnetic field strength for those simulations where the magnetic field only slows the expansion (simulations~B and C). 
However, should the field stop completely the expansion in the perpendicular direction, as is the case for simulation~D, 
then the excess energy is used to expand further in the direction parallel to the field, increasing the asymmetry of the bubble. 

In the non-magnetic case, the final bubble is spherical and has a radius of approximately 30\,pc. 
This gives the shocked wind bubble a final volume of more than 110\,000\,pc$^3$. 
For the 20\muG\, interstellar magnetic field, the final bubble is almost cylindrical, with dimensions $r\,=\,7$\,pc and $z\,=\,100$\,pc, 
for a total volume of about 15\,000\,pc$^3$. 
As a result, the density of the shocked wind is eight times higher for the magnetic model. 
This makes the radiative cooling in the shocked wind $8^2\,=\,64$ times as effective, further reducing the thermal pressure in the shocked wind. 
(The length-width ratio for the bubbles inside an interstellar magnetic field may have been exaggerated by the fact that our 2D simulations 
enforce axi-symmetry, which in 3D could be broken due to local effects.)

In the low density, warm, ISM (E), the magnetic field dominates the force on the outside of the expanding bubble (owing to a lack of either thermal pressure or inertia). 
As a result the bubble becomes far more asypherical than when a similar magnetic field is combined with high density gas (B). 
Furthermore, the low density allows the bubble to grow to a much larger size. 
In other respects, the bubble evolution still follows the same pattern as in simulations~B and C. 

For the special case where a warm, high density ISM is combined with an interstellar magnetic field (simulation~F), the evolution of the circumstellar bubble 
deviates radically from the established norm; particularly during the red supergiant phase, when the hot, shocked wind bubble 
disappears completely, to be replaced with a purely radiative shock between the red supergiant wind and the ISM.

\subsection{From circumstellar bubbles to super bubbles}
Although we limit ourselves to the circumstellar medium of one star, in reality massive stars are formed in clusters. 
This changes the evolution of the circumstellar medium, because the wind-blown bubbles of the individual stars 
encounter one another as they expand and merge to form a single super bubble as shown by \citet{vanMarleetal:2012a} and \citet{Krauseetal:2013b,Krauseetal:2013a}. 
As demonstrated both in this paper, as well as in \citet{vanMarleetal:2012a}, each circumstellar bubble can effectively contain the expansion of the supernova, 
thus even when the stars explode no interaction occurs, unless the bubbles have already merged during the wind-driven expansion phase. 
\citet{Tomisaka:1990,Tomisaka:1992} and \citet{Ferriere:1991} explored the influence of the interstellar field on the outer shape of a super bubble, 
but could not model the mergers, because they approximated the cluster as a point source of kinetic energy.  

The presence of a strong interstellar magnetic field could potentially prevent such mergers, depending on the number of stars, 
their distribution over the local space and the strength of the magnetic field. 
Along the field lines, the individual bubbles will merge easily. 
However, perpendicular to the field lines, the bubbles will have to overcome the interstellar field, the strength of which is enhanced because
the field is compressed from multiple directions at the same time. 
The assumption of a large scale, uniform magnetic field  would no longer be correct under these circumstances. 
However, such a field is not necessary to influence the shape of the circumstellar bubbles. 
In a dense cluster, the local ISM density at the start of star formation will most likely be much higher than 10\,cm$^{-3}$. 
At the same time, the magnetic field is likely to be at least an order of magnitude stronger \citep{Crutcheretal:1999}. 
As long as the field is uniform on the same scale as the individual bubbles (a few parsec for ISM densities of molecular clouds) 
the expansion of the bubbles can be contained. 
The final outcome would depend on the number and density of the star cluster. 
If the individual bubbles are kept from merging, this would result in a filamentary structure, not only in terms of density, but also in terms of metallicity, 
because the enriched supernova material would effectively be trapped inside the elongated wind bubbles and only merge with the ISM when 
the bubbles collapsed due to a lack of internal pressure after the deaths of their progenitor stars. 
This complex investigation will be carried out in further works.

The general shape of superbubbles can serve as a test for the influence of the interstellar magnetic field on the expansion of wind-driven bubbles, 
even if the field does not prevent mergers. 
According to \citet{Heiles:1979} super bubbles tend to be elongated along the galactic plane. 
\citet{Ferriere:1991} argued that this indicated that magnetic fields, 
rather than density stratification dominated the shape of super bubbles, because the magnetic field orientation in the galaxy tends 
to be aligned with the disk, whereas density stratification would increase the expansion perpendicular to the disk. 
A similar alignment was observed by \citet{WeidmannDiaz:2008} for planetary nebulae. 

Besides bubble dynamics and morphological aspects, the interplay of bubbles and their turbulent environment is essential to understand the propagation of 
cosmic rays \citep{Bykov:2001,FerrandMarcowith:2010}. 
Cosmic rays, depending on the way they are transported inside these structures, may also impact the bubble dynamics; a possibility that has been largely overlooked until now. 
These complex aspects deserve a full 3D approach including the impact of energetic particles. 
Software capable of handling such a model is currently under development.

\subsection{Effects on the shape of shells inside the bubble}
Although the field cannot affect the events inside the bubble, such as the formation of circumstellar nebulae and supernova remnants, 
 directly, it can affect them indirectly, because these events are constricted by the size and shape of the bubble. 

For the weak (5\muG) and intermediate (10\muG) fields the shells formed inside the bubble as a result of changing wind parameters 
are largely unaffected by the presence of the interstellar magnetic field. 
The cross-section of the bubble, even along the minor axis, is large enough to allow the shells to form and evolve as though the whole bubble were spherical. 
For the strong (20\muG) magnetic field the situation is different. 
In this case the bubble has become so constricted by the magnetic field, that the evolution of the WR shell is 
impacted at an early stage and forced into an asymmetric shape 
that resembles a jet, even though it was originally formed as a spherical shell. 
The same holds true for the shells formed by the expansion of supernova remnants. 

The influence of the interstellar magnetic field on the shape of
supernova remnants can potentially serve as a means of testing the model observationally. 
The interstellar magnetic field cannot influence the shape of an expanding supernova blastwave directly \citep{Manchester:1987}. 
Therefore, any correlation between the shape of supernova remnants and the morphology of the interstellar magnetic field has to be caused indirectly:  
The magnetic field shapes the wind cavity into which the supernova explodes and thereby influences the shape of the supernova remnant indirectly. 
 That the wind-cavity influences the shape of the supernova remnant was first demonstrated through 2D simulations by \citet{Rozyczkaetal:1993}. 
\citet{Arnal:1992} proposed ambient magnetic fields as an explanation for barrel shaped cavities around Wolf-Rayet stars, which in turn could 
be the source of asymmetry in supernova remnants and 
\citet{Gaensler:1998} showed that galactic supernova remnants tend to be aligned with the galactic disk, which seems to confirm that the 
interstellar magnetic field affects their expansion. 
A drawback of this test is that an asymmetric cavity around a Wolf-Rayet star (and as a result an asymmetric supernova remnant) 
can in theory be formed through the motion of its progenitor star relative to the surrounding ISM. 
If the motion is supersonic, it causes the stellar wind to form a bowshock structure, rather than a normal bubble. 
Although no proven case of a bow shock around a Wolf-Rayet star has been observed, 
it can theoretically account for the existence of asymmetric supernova remnants \citep{Gvaramadze:2006,Meyer:2015}.

 In simulation~E the a-spherical shape of the bubble is largely irrelevant to the development of the internal shells, on account of its shear size. 
Even the Wolf-Rayet ring nebula has broken up and largely dissipated before it collides with the outer edge of the bubble. 

Again, simulation~F, with its unique bubble evolution presents a special case. 
Because the main sequence wind bubble disappears completely, there is no separate red supergiant shell. 
Instead, the red supergiant wind impacts directly on the ISM. 
A Wolf-Rayet ring nebula is formed inside the red supergiant wind, 
but does not survive the collision with the outer shell.

\subsection{Lack of instabilities}
In all three simulations including a magnetic field (B, C and D) the magnetic field effectively prevents the formation of instabilities in the outer shell. 
The magnetic pressure counteracts the compression of the shell, preventing the formation of thin-shell instabilities. 
In theory, Rayleigh-Taylor instabilities could still form \citep{Breitschwerdtetal:2000,Stonegardiner:2007} 
and we see some sign of the onset of such instabilities along the contact discontinuity between shocked wind and shocked ISM 
(see centre panels of Figs.~\ref{fig:wsnB10_fig1} and \ref{fig:wsnB20_fig1}). 
 Rayleigh-Taylor instabilities are also evident in Simulation~E when the interstellar gas at the contact discontinuity is temporarily compressed 
by the impact of the supernova remnant (right panel of Fig.~\ref{fig:wsnB05lowdens_fig2}).
However, further growth of such instabilities would involve motion of the gas perpendicular to the field lines 
and the forces creating the instability are insufficient to overcome the magnetic field. 
This side-effect may be of interest in other circumstances as well and can, 
for example, explain the absence of instabilities in the \emph{Herschel} observations of the bow-shock of \textalpha-Orionis
\citep{Decinetal:2012,vanMarleetal:2014}. 
When studying the effect of the interstellar magnetic field on the instabilities, one should keep in mind, that our models are limited 
by their 2.5-D nature. 
The interaction between a magnetic field and hydrodynamical instabilities is an inherently 3D problem \citep{Breitschwerdtetal:2000,Stonegardiner:2007} 
and should be further investigated before conclusions can be drawn 
(see the discussions in \citet{Arthuretal:2011,KimKim:2014}).

\subsection{Implications for gamma-ray burst progenitors}
Because the magnetic field keeps the bubble confined, the thermal pressure of the shocked wind remains higher, 
which places location of the wind termination shock, the balance between thermal pressure and ram-pressure, closer to the star, 
though not as high as predicted by the analytical approximation (See Sect.~\ref{sec-ISM}).
This can be of interest in the case of possible long gamma-ray burst (GRB) progenitors. 
The afterglows of many long GRBs show evidence of a constant density, rather than wind-like environment, 
despite the fact that these bursts are thought to originate from massive stars \citep[e.g.][]{Chevalieretal:2004}. 
This has been explained as being the shocked wind, rather than the free-streaming wind. 
However, this would indicate that the wind termination shock is very close ($\ll\,0.1$\,pc) to the star. 
This has been explained in several ways, including a high ISM density \citep{eldridgeetal:2006,vanMarleetal:2006}, 
and/or a high ISM temperature \citep{Chevalieretal:2004,vanMarleetal:2006}. 
However, it is possible that interstellar magnetic fields play a role here as well. 
Although the magnetic fields applied in this paper do not reduce the radius of the wind termination shock 
as much as is required to explain the GRB afterglow observations, 
higher magnetic field strengths have been observed in molecular clouds \citep{Crutcheretal:1999,Crutcher:2012}. 
These fields, combined with other factors such as high ISM density and temperature, and a reduced mass loss rate at low metallicity, 
could, in theory, reduce the termination shock radius to less than 0.1\,pc. 

\begin{figure*}
\FIG{
 \centering
\mbox{
\subfigure
{\includegraphics[width=0.5\textwidth]{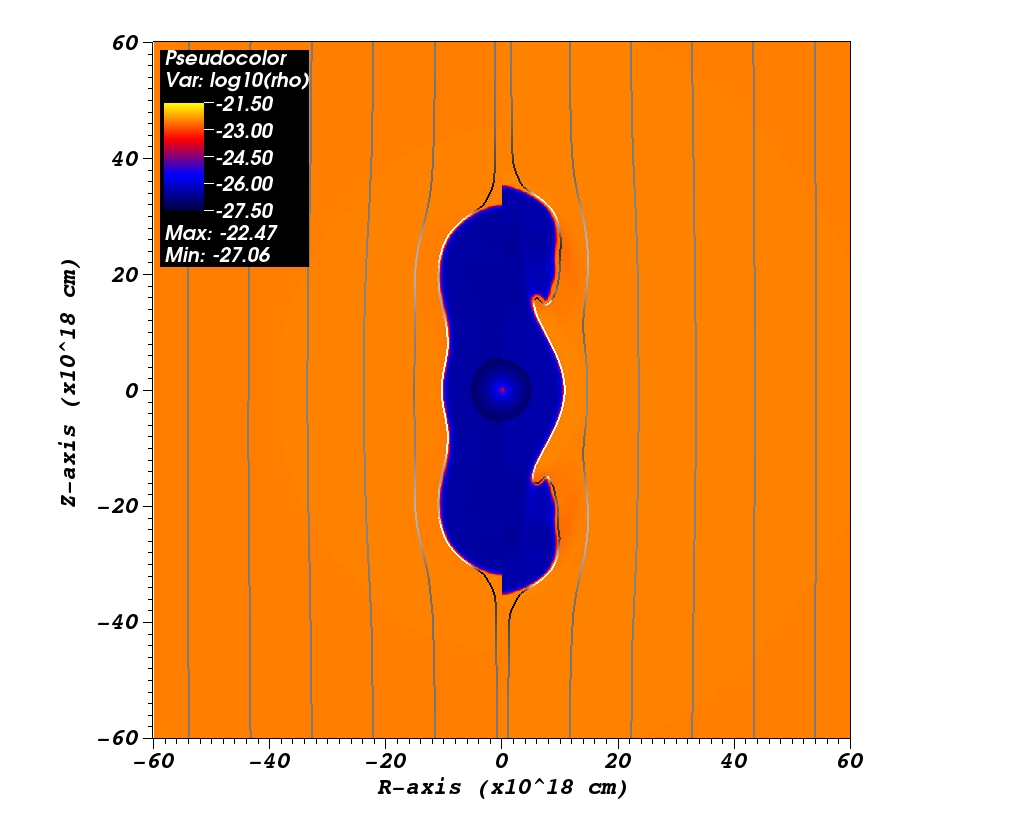}}
\subfigure
{\includegraphics[width=0.5\textwidth]{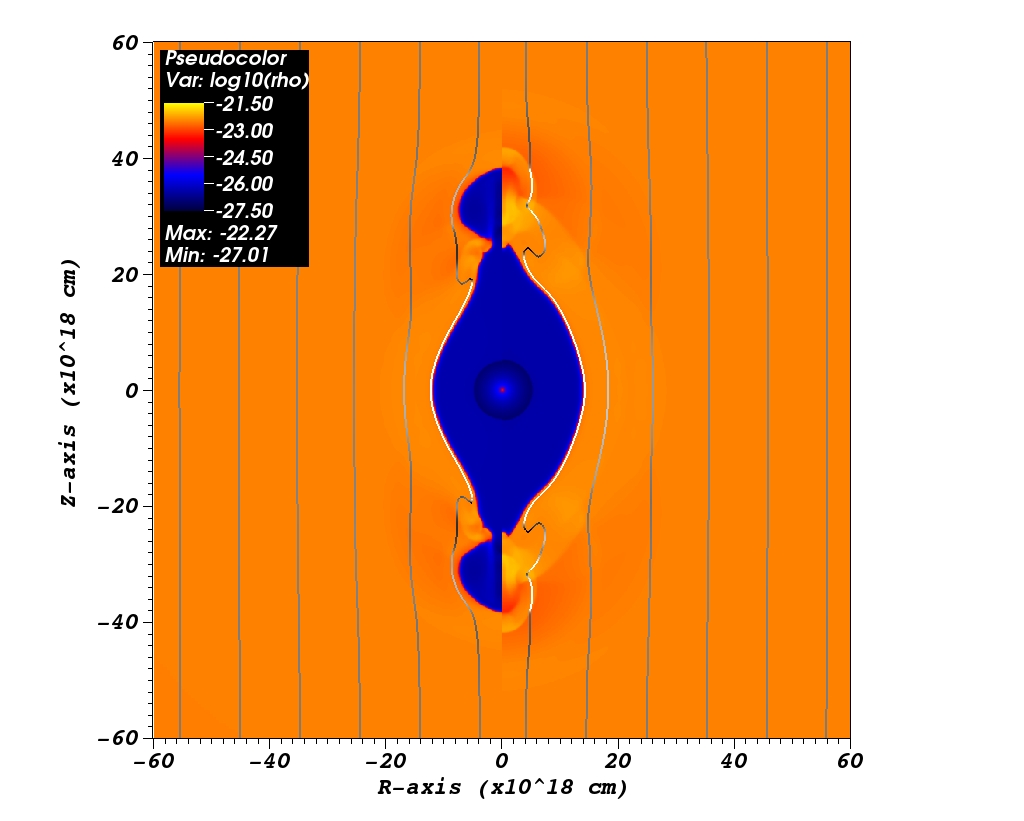}}}
}
\caption{ The shape of the bubble for simulation~F after 3, 3.5, 3.75, and 4\,Myr. 
The tips of the bubble are squeezed off by the magnetic field. This effect is enhanced by the 2D nature of our simulations and may not actually occur in 3D.
}
\label{fig:wsnB5warm_2Dissue}
\end{figure*}

\subsection{Limitations of the model}
\label{sec-limitations}
We are aware that the ISM parameter range explored in this work is limited. 
The environment of massive stars is very complex, inhomogeneous, and intermittent. 
The structure of the parent molecular cloud includes media of different properties from diffuse to dense cores. 
The molecular clouds are known to be turbulent
(see a recent review by \citet{HennebelleFalgarone:2012}), pervaded by magnetic fields, the strength of which can exceed 20\muG\, by an order of magnitude \citep{Crutcher:2012}.
The turbulent motions in the interstellar medium are known to have an impact on the shape and the dynamics of the stellar bubble \citep{Korpietal:1999,SilichFranco:1999}. 
H$_{\rm II}$ regions also harbor turbulent motions showing particular statistical properties \citep{Medinaetal:2014}; turbulent motions associated with various processes as: 
internal globules evaporation and/or interaction with stellar winds. 

More realistic models of the ISM into which the circumstellar bubbles of massive stars are blown have to include the combination of magnetic field effects in turbulent molecular clouds and/or in H$_{\rm II}$ regions. 
Such models deserve a full 3D treatment. 
Furthermore, before such comprehensive 3D models are attempted, it is necessary to first explore how each of the many interacting 
forces in the ISM affects the evolution of a circumstellar bubble. 
To this end we have deliberately kept the gas properties of the ISM a constant in order to isolate the influence of the large scale 
magnetic field on the evolution of the circumstellar bubbles. 

The 2D nature of our models, while greatly reducing computation time, forms a second limitation. 
For purely hydrodynamic interactions \citet{vanMarleKeppens:2012} showed that instabilities in Wolf-Rayet ring nebulae tend to grow more quickly in 3D models than in 2D, 
though the difference tends to decrease over time. 
As to the effect of 2D symmetry on the magneto-hydrodynamical interaction between the bubble and the interstellar magnetic field, this is a more complicated question. 
\citet{vanMarleetal:2014b} showed that a 3D model of the interaction between an AGB wind and the interstellar magnetic field yields 
the same qualitative and quantitative result as a 2D model. 
However, that model lacked some of the features that can occur in the models presented here: in particular the phenomenon observed in 
simulations~D and F, where part of the bubble is separated from the rest because of the magnetic field pressure. 
 This effect is particularly visible in simulation~F  during the late main sequence (Fig.~\ref{fig:wsnB5warm_2Dissue}). 
As the bubble expands the thermal pressure in the regions far from the star decreases to the point where it becomes less than the combined magnetic and thermal pressure
 of the ISM. 
As a result these regions are cut off from the rest of the bubble and disappear since they no longer have a stellar wind to supply them with internal energy. 
In a 3D model, which has an additional degree of freedom for expansion,  this effect may be reduced, or even disappear completely. 
If it occured at all, it would not show the axial symmetry that occurs in 2D. 
The ``eye-like''  shape of the bubble after the tips have been reduced is not a 2D artefact and has been observed in 3D simulations of AGB winds \citet{vanMarleetal:2014b}

 So far we have ignored the existence of stellar magnetic fields. 
Simulations of the circumstellar medium of low mas stars \citep[e.g.][]{RozyczkaFranco:1996,Garcia-Seguraetal:1999} 
show that toroidal stellar magnetic fields of AGB and post-AGB stars can create an extraordinary variation in the shapes of planetary nebulae by changing the direction of the stellar wind, 
which for these kind of stars tend to be weak.  
Simulations of the interaction between the stellar wind and magnetic field of high mass stars tend to focus on dipole fields in a region close to the star 
(primarily the wind acceleration zone) rather than the extended wind cavity. 
Such simulations \citep[e.g.][]{uddoulaowocki:2002,Townsendetal:2007,uddoulaetal:2008,uddoulaetal:2009} show that while a dipole stellar magnetic field 
can influence the morphology of the stellar wind close to the star, at longer distances the stellar wind, 
which is faster and far more powerful than the winds of low mass stars, starts to dominate, 
forcing the field lines to align with the expanding wind to form a magnetic monopole. 
This allows the wind to expand spherically as though the magnetic field no longer exists, leaving the magnetic field largely irrelevant to the 
morphology of the circumstellar shells. 
Exactly how such a magnetic field would interact with the interstellar magnetic field is difficult to predict, but it should be noted that the 
distance from the star would greatly reduce the stellar magnetic field strength. 
For example, \citet{Wadeetal:2012} estimate the surface dipole strength of \object{NGC~1624-2}, the strongest magnetic field ever measured for an O-star, 
to be at least 20\,kG. 
Assuming that the field strength decreases with the distance squared (as would be the case for a monopole) and a stellar radius of about $11\rso$ \citep{Wadeetal:2012} 
this would reduce the field to approximately $1.2\times10^{-3}$\,\muG\, at a distance of one parsec, 
significantly weaker than even the weak interstellar fields in the galactic halo. 
For more representative O-star magnetic field strengths of 10-1000\,G \citet[][and references therein]{Walderetal:2012} 
the effective field strength at the interaction point with the ISM would be even weaker. 
Note that if the stellar magnetic field were to retain its dipole (or higher order) configuration despite the stellar wind, its field strength would decrease even faster. 
Considering that wind-cavities tend to be on a multi-parsec scale, it seems reasonable to assume that the interaction between the stellar and 
interstellar magnetic fields would only be relevant for very young stellar objects 
and become irrelevant once the wind cavity is formed. 
However, these simulations only involved hot stars such as O- and B-stars. 
During the red supergiant phase the ram pressure of the wind is much smaller and the wind velocity is lower, reminiscent of an AGB star. 
For such as star a toroidal magnetic field, if present, may influence the direction of the wind in a manner similar to that predicted for AGB stars. 
However, the scale of the circumstellar cavity (at least an order of magnitude large than those of low mass stars) would still limit any direct interaction between 
the stellar magnetic field and the ISM field.

\section{Conclusion}
\label{sec-conclusions}
The shape, size, and evolution of circumstellar bubbles are influenced by many factors, such as the properties of the stellar wind, 
the motion of the star relative to its surrounding,  the density and temperature of the interstellar gas, and 
the strength and shape of the interstellar magnetic field.
In this paper we have focussed our investigation on the influence of the strength of the interstellar magnetic field on the shape and size of the 
circumstellar bubble around a massive star.

We have limited our calculations to regular magnetic fields of different strengths that are characteristic of different regions in the galaxy: 
from low magnetic found in the disk to higher values found in the galactic bulge. 
The simulations are 3D axi-symmetric and permit a fast exploration of the parameter space. 
Although simple with respect to any realistic interstellar medium, the simulation setups allowed us to derive some important conclusions.  
As expected, the interstellar magnetic field constrains the expansion of a circumstellar bubble in the direction perpendicular to the field. 
If the magnetic field is strong enough to generate a pressure that exceeds the ram-pressure of the stellar wind at the wind termination shock, 
the expansion in the perpendicular direction stops completely. 
This leads to the formation of a strongly asymmetric bubble, the size of which is so constrained 
that it affects the evolution of temporary circumstellar features such as Wolf-Rayet ring nebulae. 
Even for the powerful wind of an O-type massive star, galactic magnetic fields can be strong enough to do so. 

In the future we will extend these simulations to 3D in order to include random fluctuations in the magnetic field. 
These fluctuations can be of similar strength to the uniform field component \citep{Beck:2009}. 
Such 3D models will also allow us to further investigate the influence of the magnetic field on the local instabilities 
in the swept-up shell. These 3D simulations will be complemented by a setup that includes the detailed structure of the
star bubble environment that comprises the parental molecular clouds and H$_{\rm II}$ regions. 

\begin{acknowledgements} 
A.J.v.M.\ acknowledges support from FWO, grant G.0277.08, K.U.Leuven GOA/2008/04 and GOA/2009/09. 
Z. Meliani acknowledges  financial support from the PNHE. 
\end{acknowledgements}

\bibliographystyle{aa}
\bibliography{vanmarle_biblio}

\IfFileExists{vanmarle_biblio.bbl}{}
 {\typeout{}
  \typeout{******************************************}
  \typeout{** Please run "bibtex \jobname" to obtain}
  \typeout{** the bibliography and then re-run LaTeX}
  \typeout{** twice to fix the references!}
  \typeout{******************************************}
  \typeout{}
 }
\listofobjects
 
 \Online
\begin{appendix}  
\section{Animations of the evolution of circumstellar bubbles}
\label{sec-animations}
This appendix contains the results of all our simulations in animation form. 
Keep in mind that the frame rate differs between the stellar evolution and supernova phases because of the different timescales involved.

\begin{figure}
\caption{Animation A\_wind, shows the evolution of the circumstellar bubble according to simulation~A (no interstellar magnetic field) 
for the pre-supernova phase.}
\end{figure}

\begin{figure}
\caption{Animation A\_sn shows the expansion of the supernova remnant for simulation~A.} 
\end{figure}

\begin{figure}
\caption{Animation B\_wind shows the pre-supernova evolution of the circumstellar bubble according to simulation~B ($B_{\rm ISM}=5\,$\textmu G).}
\end{figure}

\begin{figure}
\caption{Animation B\_sn shows the expansion of the supernova remnant for simulation~B.}
\end{figure}

\begin{figure}
\caption{Animation C\_wind shows the pre-supernova evolution of the circumstellar bubble according to simulation~C ($B_{\rm ISM}=10\,$\textmu G).}
\end{figure}

\begin{figure}
\caption{Animation C\_sn shows the expansion of the supernova remnant for simulation~C.}
\end{figure}

\begin{figure}
\caption{Animation D\_wind shows the pre-supernova evolution of the circumstellar bubble according to simulation~D ($B_{\rm ISM}=20\,$\textmu G).}
\end{figure}

\begin{figure}
\caption{Animation D\_sn shows the expansion of the supernova remnant for simulation~D.}
\end{figure}

\begin{figure}
\caption{Animation E\_wind shows the pre-supernova evolution of the circumstellar bubble according to simulation~E ($B_{\rm ISM}=5\,$\textmu G, 
low density warm ISM).}
\end{figure}

\begin{figure}
\caption{Animation E\_sn shows the expansion of the supernova remnant for simulation~E.}
\end{figure}

\begin{figure}
\caption{Animation F\_wind shows the pre-supernova evolution of the circumstellar bubble according to simulation~F ($B_{\rm ISM}=5\,$\textmu G, high density warm ISM).}
\end{figure}

\begin{figure}
\caption{Animation F\_sn shows the expansion of the supernova remnant for simulation~F.}
\end{figure}

\end{appendix}

\end{document}